\documentclass[12pt]{article}
\pdfoutput=1
\usepackage{amsmath}
\usepackage{amssymb}
\usepackage{epsfig}
\usepackage{cite}
\usepackage{fontenc}
\usepackage{graphicx}
\usepackage{float}
\usepackage[lofdepth,lotdepth,caption=false]{subfig}
\usepackage{color}
\usepackage{booktabs}
\usepackage{tabularx}
\usepackage{multirow}
\usepackage{longtable}
\usepackage{lscape}
\usepackage{dcolumn}
\usepackage{rotating}
\usepackage[normalem]{ulem}
\definecolor{darkgreen}{cmyk}{1,0,1,0.4}
\definecolor{brown}{cmyk}{0,0.8,1,0.2}
\definecolor{darkred}{cmyk}{0,1,1,0.2}

\allowdisplaybreaks[4] % for page break
\makeatletter
\renewcommand{\fnum@table}{\textbf{\tablename~\thetable}}
\renewcommand{\fnum@figure}{\textbf{\figurename~\thefigure}}
\makeatother

\newcounter{myenumi}

\renewcommand{\themyenumi}{\roman{myenumi}}
{\end{list}}

\newlength{\myem}
\newcounter{mysubequation}[equation]

\makeatletter
\renewcommand{\section}{\@startsection{section}{1}{0em}{-\baselineskip}%
{\baselineskip}{\normalfont\large\bfseries}}
\renewcommand{\subsection}%
{\@startsection{subsection}{2}{0em}{-0.7\baselineskip}%
{0.7\baselineskip}{\normalfont\bfseries}}
\makeatother

\newcommand{\bi}{\begin{itemize}}
\newcommand{\ei}{\end{itemize}}

\def\beq{\begin{equation}}
\def\eeq{\end{equation}}

\newcommand{\bea}{\begin{eqnarray}}
\newcommand{\eea}{\end{eqnarray}}

\newcommand{\ie}{{\it i.e.}}

\newcommand{\eet}{\varepsilon_{e\tau}}

\newcommand{\eem}{\varepsilon_{e\mu}}

\newcommand{\eeta}{|\varepsilon_{e\tau}|}

\newcommand{\eema}{|\varepsilon_{e\mu}|}

\newcommand{\nue}{\ensuremath{\nu_e}}
\newcommand{\numu}{\ensuremath{\nu_\mu}}

\def\epsilon{\varepsilon}
\def\eps{\varepsilon}

\newcommand\schd{Schr$\ddot{\rm o}$dinger}

\def\<{\langle}
\def\>{\rangle}

\def\dfrac#1#2{{\displaystyle\frac{#1}{#2}}}

\def\lsim{\mathrel{\rlap{\lower4pt\hbox{\hskip1pt$\sim$}}
    \raise1pt\hbox{$<$}}}         %less than or approx. symbol
\def\gsim{\mathrel{\rlap{\lower4pt\hbox{\hskip1pt$\sim$}}
    \raise1pt\hbox{$>$}}}         %greater than or approx. symbol

\newcommand{\pmue}[1]{\ensuremath{{ P}_{\mu e} \left(#1 \right)}}
\newcommand{\pmueb}[1]{\ensuremath{\bar {P}^{}_{\mu e} \left(#1 \right)}}

\newcommand{\delpcp}[1]{\ensuremath{\Delta {P}_{\mu e}  \left(#1 \right)}}

\newcommand{\sumpcp}[1]{\ensuremath{\sum {P}_{\mu e}  \left(#1 \right)}}

\newcommand{\acp}[1]{\ensuremath{{A}^{CP}_{\mu e} \left(#1 \right)}}

\begin{document}
%
%%%%%%%%%%%%%%%%%%%%%%%%%%%%%%%%%%%%%%%%%%%%%%%%%%%%%%%%%%%%%%%%%%%%%
%%%%                     Title-page                              %%%%
%%%%%%%%%%%%%%%%%%%%%%%%%%%%%%%%%%%%%%%%%%%%%%%%%%%%%%%%%%%%%%%%%%%%%

\begin{titlepage}

\renewcommand{\thefootnote}{\alph{footnote}}

\vspace*{-3.cm}
\begin{flushright}
% Report numbers

\end{flushright}

%\vspace*{0.5cm}

\renewcommand{\thefootnote}{\fnsymbol{footnote}}
\setcounter{footnote}{-1}

{\begin{center}
{\large\bf Probing CP violation signal at DUNE in presence of non-standard neutrino interactions 
\\[0.2cm]
}
\end{center}}

\renewcommand{\thefootnote}{\alph{footnote}}

\vspace*{.8cm}
\vspace*{.3cm}
{\begin{center} 
            {{\sf 
                Mehedi Masud$^\star$~\footnote[1]{\makebox[1.cm]{Email:}
                masud@hri.res.in},
				Animesh Chatterjee$^\dagger$~\footnote[2]{\makebox[1.cm]{Email:}
                animesh.chatterjee@uta.edu}, and 
                Poonam Mehta$^{\ddagger}$~\footnote[3]{\makebox[1.cm]{Email:}
                pm@jnu.ac.in}

               }}

\end{center}}
\vspace*{0cm}
{\it 
\begin{center}
\footnotemark[1]%
$^\star$\, Harish-Chandra Research Institute, Chhatnag Road, Jhunsi, Allahabad 211 019, India

\footnotemark[2]%
$^\dagger$\, Department of Physics, University of Texas at Arlington,
Arlington, TX 76019, USA 

\footnotemark[3]%
$^\ddagger$ \,     School of Physical Sciences,       Jawaharlal Nehru University, 
      New Delhi 110067, India

\end{center}}

\vspace*{1.5cm}

%\begin{center}
%{\Large \today}
%\end{center}

{\Large 
\bf
\begin{center} Abstract 
\end{center} 
 }

We discuss the impact of  non-standard neutrino matter interactions (NSI) in propagation on the determination of CP phase in the context of the long baseline accelerator experiments such as Deep Underground Neutrino Experiment (DUNE).
 DUNE will mainly address the issue of CP violation in the leptonic sector.  Here we study the role of NSI and its impact on the question of observing the CP violation signal at DUNE. We consider two scenarios of oscillation with three active neutrinos in absence and presence of NSI. We elucidate the importance of ruling out subdominant new physics 
effects introduced by NSI in inferring CP violation signal at DUNE by considering NSI terms collectively as well as by exploiting the non-trivial interplay of moduli and phases  of the NSI terms. 
We demonstrate the existence of NSI-SI degeneracies 
 which need to be eliminated in reliable manner in order to make conclusive 
 statements about the CP phase.   
\vspace*{.5cm}

\end{titlepage}

\newpage

\renewcommand{\thefootnote}{\arabic{footnote}}
\setcounter{footnote}{0}

%----------------------------------------------------------------------%
\section{Introduction}

With the discovery of Higgs boson at the Large Hadron Collider being recognized with the Nobel 
prize in 2013~\cite{nobel2013}, the Standard Model (SM) proposed by Glashow, Weinberg 
and Salam in 1967 has gained full acceptance. 
However there are compelling reasons hinting towards physics beyond the scale of the SM.  
 A series of neutrino oscillation experiments have established the phenomenon of neutrino flavour oscillations due to non-zero masses which requires new physics as neutrinos are massless in the SM.  The latest 
   Nobel prize has been awarded for this significant discovery~\cite{nobel2015}. 
  Several exciting ideas and directions have been proposed which lead to the physics beyond  the SM.  
   Out of the nine flavour parameters in the standard three flavour mixing framework, only six~\footnote{The absolute  mass scale and the two Majorana phases are not accessible in oscillation experiments.} can be accessed via oscillation experiments -  three angles ($\theta_{12},\theta_{13},\theta_{23}$), two mass squared differences ($\delta m^2_{31}, \delta m^2_{21}$) and a single Dirac-type CP\,\footnote{CP refers to charge conjugation and parity symmetry.} phase ($\delta$). The angles and the mass-squared differences (and absolute value of only one of them) have been measured with great precision, only recently it has become possible to pin down the CP phase in the leptonic sector - thanks to the measurement of $\theta_{13}$ and largeness of its value~\cite{GonzalezGarcia:2012sz,Capozzi:2013csa,Forero:2014bxa}.

 The main focus of the ongoing and future neutrino experiments is to  address the question of 
 neutrino mass hierarchy \ie, if sign $(\delta
 m^2_{31}) > 0 $ (normal hierarchy, NH) or sign $(\delta
 m^2_{31}) < 0 $ (inverted hierarchy, IH)\,\footnote{$\delta m^2 _{31} = m^2 _{3} - m^2_1$. }, measurement of 
 the CP phase ($\delta$) and establishing the correct octant
 of the mixing angle $\theta_{23}$. Some of the promising efforts include reactor experiments such as 
 Jiangmen Underground Neutrino Observatory (JUNO)~\cite{juno}, accelerator experiments~\cite{accelerator} such as Numi Off-axis electron neutrino Appearance (NOvA) experiment~\cite{nova}, Deep Underground Neutrino Experiment 
 (DUNE)\,\footnote{erstwhile the Long Baseline Neutrino Experiment (LBNE).}~\cite{2013arXiv1307.7335L,Bass:2013vcg} and atmospheric neutrino experiments such as India-based Neutrino Observatory (INO)~\cite{icalreport} and Precision IceCube Next Generation Upgrade (PINGU) experiment~\cite{pingu}.

The question of whether CP is violated in the leptonic sector is of prime importance in astrophysics, cosmology and particle physics today. Neutrino oscillations at long baselines offer a promising option to infer leptonic CP violation~\cite{Arafune:1996bt,Tanimoto:1996ky,Bilenky:1997dd,Barger:1980jm,Diwan:2003bp,Brahmachari:2003bk,Akhmedov:2004ve,Marciano:2006uc,pakvasa}. Leptonic CP violation is a possible ingredient to explain the observed baryon asymmetry of the Universe via leptogenesis (for a review, see ~\cite{Davidson:2008bu}).  Within the SM, effects due to CP violation reside in a phase of the $3 \times 3$ mixing matrix known as the Cabibbo-Kobayashi-Maskawa  (CKM) mixing matrix for  quarks~\cite{cabibbo,km,Pakvasa:1975ti,pdg2014}.  Analogously, one expects a similar mixing matrix for leptons if neutrinos are massive given by 
 $U_{PMNS} = U^{l\dagger}_L U^{\nu}_L$ ($M_\nu^{diag} = U^\nu_L M_\nu U^{\nu T} _L$ for neutrinos and $M_l^{diag} = U^l_L M_l U^{l \dagger} _L$ for the charged leptons) which is called the Pontecorvo-Maki-Nakagawa-Sakata (PMNS) matrix~\cite{pontecorvo,mns,pdg2014} and consequently, CP phase(s) appear in neutrino mixing. 
 For the three flavour case in vacuum, the only source of CP violation in mixing phenomena  is the Dirac-type CP phase, $\delta$~\cite{Bilenky:1984fg}.  This is usually referred to as the intrinsic CP phase.

 It is well-known that for baselines $\sim {\cal O} (1000)$ km, the standard Earth matter effects~\cite{Wolfenstein:1977ue,Mikheev:1987qk} are non-negligible. This poses a problem in the 
  determination of intrinsic CP phase as  matter induces additional CP violating effects in the oscillation formalism, commonly referred to as extrinsic (fake) CP violation effects~\cite{Arafune:1997hd}. Any new physics in neutrino interactions can, in principle, allow for flavour changing interactions thereby allowing for additional CP violating phases which can complicate the extraction of the intrinsic CP phase further. The high precision offered  by DUNE facilitates probing  new physics phenomenon  such as additional sterile neutrino states which has been recently studied~\cite{Berryman:2015nua,Gandhi:2015xza}, probing Lorentz and CPT violation (e.g. in~\cite{Datta2004356,Chatterjee:2014oda}) as well as NSI during propagation (e.g. in ~\cite{Ohlsson:2012kf,Blennow:2005qj,Adhikari:2012vc}) with high sensitivity.  In the present article, we explore the impact of NSI in propagation on CP violation signal at upcoming long baseline neutrino experiments.

We use DUNE~\cite{2013arXiv1307.7335L,Bass:2013vcg} as an example in the present work. It utilises a mega-watt class proton accelerator (with beam power of upto 1.2 MW) at Fermi National Accelerator Laboratory (Fermilab) to produce high intensity neutrino source. For the far detector, a massive liquid argon time-projection chamber (LArTPC) would be deployed deep underground at a depth of 4850 feet at the Sanford Underground Research Facility located at the site of the former Homestake Mine in Lead, South Dakota (where Ray Davis carried out the solar neutrino experiment during 1967-1993) and is about 1300 km from the neutrino source at Fermilab. In addition, a high precision near neutrino detector is planned at a distance of approximately 500 m from the target at Fermilab site. The baseline of 1300 km is expected to deliver optimal sensitivity to  CP violation, measurement of $\delta$ and at the same time is long enough to address the question of neutrino mass hierarchy~\cite{Barger:2013rha,Barger:2014dfa,Qian:2015waa}. It is worth mentioning that CP violation can be established at $3\sigma$  level if we consider  DUNE for at least $\sim 68\%$ of CP phase values~\cite{2013arXiv1307.7335L,Bass:2013vcg} and it has been shown that a combination of different experiments can increase this fraction to $\sim 80\%$ for reasonable exposures~\cite{Barger:2014dfa}.

The plan of the article is as follows. We first briefly outline the NSI framework and give the 
present constraints on NSI parameters in Sec.~\ref{sec:framework}. We then go on to describe 
observable CP asymmetry for the particular channel $\nu_\mu \to \nu_e$ relevant for DUNE both 
in vacuum and in matter (SI and NSI) in Sec.~\ref{sec:cpvprob}. 
We present our results and discussions in Sec.~\ref{sec:results} and discuss the
 event rates obtained at DUNE far detector in Sec.~\ref{sec:events}. 
We end with conclusions in Sec.~\ref{sec:conclude}.

%----------------------------------------------------------------------%
\section{Non-standard Neutrino Interactions} 
\label{sec:framework}

We consider effects that can be  phenomenologically described by neutral current (NC) type neutrino NSI of the form     
\begin{equation}
\label{nsilag}
{\cal L}_{NSI} = -2 \sqrt 2 G_F \epsilon_{\alpha \beta}^{f\, C} ~ [\bar \nu_\alpha \gamma^\mu P_L \nu_\beta] ~[\bar f \gamma_\mu P_C f]~,
\end{equation}
where $G_F$ is the Fermi constant, $\nu_{\alpha},\nu_{\beta}$ are
neutrinos of different flavours, and $f$ is a first generation SM
fermion ($e,u,d$)~\footnote{The flavour of the
background fermion ($f$) is preserved in the interaction in coherent interactions.}.  
The chiral projection operators are
given by $P_L = (1 - \gamma_5)/2$ and $P_C=(1\pm \gamma
_5)/2$. 
{If} the NSI {arises} at
scale $M_{NP} \gg M_{EW}$ from some higher dimensional operators (of
order six or higher),  {it
would imply} a suppression of at least $\epsilon_{\alpha\beta}^{fC}
\simeq (M_{EW}/M_{NP})^2$ (for $M_{NP} \sim 1~TeV$, we have
$\epsilon_{\alpha\beta}^{fC} \simeq 10^{-2}$ (see also~\cite{Farzan:2015doa})).   
 At the level of the underlying Lagrangian, NSI coupling
of the neutrino can be to $e,u,d$. Phenomenologically, only the incoherent sum 
of contributions from different scatterers contributes to the coherent
forward scattering of neutrinos on matter. If we normalize to $n_e$, the
effective NSI parameter relevant for neutral Earth matter is 
\bea
\epsilon_{\alpha\beta} &=& \sum_{f=e,u,d} \dfrac{n_f}{n_e}
\epsilon_{\alpha\beta}^f = \epsilon_{\alpha\beta}^e +2
\epsilon_{\alpha\beta}^u + \epsilon_{\alpha\beta}^d + \dfrac{n_n}{n_e}
(2\epsilon_{\alpha\beta}^d + \epsilon_{\alpha\beta}^u) = \epsilon
^e_{\alpha\beta} + 3 \epsilon^u_{\alpha \beta} + 3
\epsilon^d_{\alpha\beta} \ ,
      \label{eps_combin}
\eea 
where $n_f$ is the density of fermion $f$ in medium crossed by the
neutrino and $n$ refers to neutrons.  Also, only the vector sum of NSI terms, $\epsilon_{\alpha\beta}^f=
\epsilon_{\alpha\beta}^{fL} + \epsilon_{\alpha\beta}^{fR}$ appears in the oscillation formalism.

%----------------------------------------------------------------------
%
In presence of NSI, the propagation of neutrinos is governed by a \schd-type equation with the effective Hamiltoninan  
\bea 
{\mathcal
H}^{}_{\mathrm{}} &=& {\mathcal
H}^{}_{\mathrm{vac}} + {\mathcal
H}^{}_{\mathrm{SI}} + {\mathcal
H}^{}_{\mathrm{NSI}} \ ,
\eea 
where ${\mathcal H}^{}_{\mathrm{vac}} $ is the vacuum Hamiltonian and
${\mathcal H}^{}_{\mathrm{SI}}, {\mathcal H}^{}_{\mathrm{NSI}}$ are
the effective Hamiltonians in presence of 
{SI alone and NSI} respectively. {Thus,}
 \bea
 \label{hexpand} 
 {\mathcal
H}^{}_{\mathrm{}} &=& 
\dfrac{1}{2 E} \left\{ {\mathcal U} \left(
\begin{array}{ccc}
0   &  &  \\  &  \delta m^2_{21} &   \\ 
 &  & \delta m^2_{31} \\
\end{array} 
\right) {\mathcal U}^\dagger + 
 {A (x)}   \left(
\begin{array}{ccc}
1+ \epsilon_{ee}  & \epsilon_{e \mu}  & 
\epsilon_{e \tau}  \\ {\epsilon_{e\mu} }^ \star & 
\epsilon_{\mu \mu} &   \epsilon_{\mu \tau} \\ 
{\epsilon_{e \tau}}^\star & {\epsilon_{\mu \tau}}^\star 
& \epsilon_{\tau \tau}\\
\end{array} 
\right) \right\}  \ ,
 \eea 
where $A (x)= 2 \sqrt{2}  	E G_F n_e (x)$ is the standard CC potential due to
the coherent forward scattering of neutrinos and $n_e$ is the electron
number density and ${\epsilon}_{\alpha \beta} \, (\equiv |\epsilon _{\alpha \beta}|
\, e^{i \phi_{\alpha\beta}})$ are complex NSI parameters.
The three flavour neutrino mixing matrix ${\mathcal
  U}$ [$\equiv {\cal U}_{23} \, {\cal W}_{13} \,{\cal U}_{12}$ with
  ${\cal W}_{13} = {\cal U}_\delta~ {\cal U}_{13}~ {\cal
    U}_\delta^\dagger$ and ${\cal U}_\delta = {\mathrm{diag}}
  \{1,1,\exp{(i \delta)}\}$] is characterized by three angles and a
single (Dirac) phase~\footnote{In the general case of n flavors the leptonic mixing matrix 
$U_{\alpha i}$ depends on $(n-1)(n-2)/2$ Dirac-type 
CP-violating  phases. If the neutrinos are Majorana particles, there are $(n-1)$ additional, so called Majorana-type CP-violating phases. } and, in the Pontecorvo-Maki-Nakagawa-Sakata (PMNS) parameterization~\cite{Beringer:1900zz}, we
have
\bea
{\mathcal U}^{} &=& \left(
\begin{array}{ccc}
1   & 0 & 0 \\  0 & c_{23}  & s_{23}   \\ 
 0 & -s_{23} & c_{23} \\
\end{array} 
\right)   
  \left(
\begin{array}{ccc}
c_{13}  &  0 &  s_{13} e^{- i \delta}\\ 0 & 1   &  0 \\ 
-s_{13} e^{i \delta} & 0 & c_{13} \\
\end{array} 
\right)  \left(
\begin{array}{ccc}
c_{12}  & s_{12} & 0 \\ 
-s_{12} & c_{12} &  0 \\ 0 &  0 & 1  \\ 
\end{array} 
\right)  \ ,
\label{u}
 \eea 
where $s_{ij}=\sin {\theta_{ij}}, c_{ij}=\cos \theta_{ij}$.   
The two additional Majorana phases in the three flavour case play no role in neutrino
oscillations and hence are not explicitly mentioned in Eq.~\ref{u}.  We define the following ratios
\begin{equation}
\lambda \equiv \frac{\delta m^2_{31}}{2 E}  \quad \quad ; \quad \quad
r_{\lambda} \equiv \frac{\delta m^2_{21}}{\delta m^2_{31}} \quad \quad ; 
\quad \quad r_{A} \equiv \frac{A (x)}{\delta m^2_{31}} \ .
\label{dimless}
\end{equation}
 For atmospheric and long
baseline neutrinos, $\lambda L \simeq {\cal O} (1)$ holds and $r_A L
\sim {\cal O} (1)$ for a large range of 
{the $E$ and $L$ values considered here}.

Let us now briefly mention the constraints  imposed on the NC NSI parameters (for more details, see \cite{Chatterjee:2014gxa}).  With the assumption that the
errors on individual NSI terms are uncorrelated, model-independent bounds on effective NC NSI terms  
\begin{equation}
\epsilon_{\alpha\beta} \lsim \left\{  \sum_{C=L,R} [ (\epsilon_{\alpha \beta}^{e C} )^2 + (3 \epsilon_{\alpha\beta}^{u C})^2 + (3 \epsilon _{\alpha \beta}^{d C})^2 ] \right\}^{1/2} \ ,\nonumber
\end{equation}
were obtained~\cite{Biggio:2009nt} which  leads to
 \begin{eqnarray}
 |\eps_{\alpha\beta}|
 \;<\;
  \left( \begin{array}{ccc}
4.2  &
0.33 & 
3.0 \\
0.33 &0.068 & 0.33 \\
3.0  &
0.33 &
21 \\
  \end{array} \right) \ ,\label{largensi}
\end{eqnarray} 
for neutral Earth matter. There are also experiments which have used the neutrino data to constrain NSI parameters. The SK NSI search in atmospheric neutrinos crossing the Earth found no evidence in favour of NSI and the study led to upper bounds on NSI
parameters~\cite{Mitsuka:2011ty} given by $|\epsilon_{\mu\tau}| < 0.033, |
\epsilon_{\tau\tau} - \epsilon_{\mu\mu} | < 0.147 $ (at 90\% CL) in a
two flavour hybrid model~\cite{Ohlsson:2012kf}\footnote{The SK
collaboration uses a different normalization ($n_d$) while writing the
effective NSI parameter (see Eq.~(\ref{eps_combin})) and hence we need
to multiply the bounds mentioned in Ref.~\cite{Mitsuka:2011ty} by a
factor of 3.}.  The off-diagonal NSI parameter
$\epsilon_{\mu\tau}$ is constrained $-0.20 < \epsilon_{\mu\tau} <
0.07$ (at 90\% CL) from MINOS data in the framework of two flavour
neutrino oscillations~\cite{Adamson:2013ovz,Kopp:2010qt}.  

We will be interested in particular channels $\nu_\mu \to \nu _e$ (and the CP transformed channel, $\bar\nu_\mu \to \bar\nu_e$) where only two of the NSI parameters ($\eem, \eet$) appear in the second order expression. Taking into account the constraints from neutrino experiments, we can write (see also~\cite{Choubey:2015xha})
\begin{eqnarray}
 |\eps_{\alpha\beta}|
 \;<\;
  \left( \begin{array}{ccc}
4.2  &
0.3 & 
0.5 \\
0.3 & 0.068 & 0.04 \\
0.5  &
0.04 &
0.15 \\
  \end{array} \right) \ .\label{tinynsi}
\end{eqnarray} 
We consider $|\eem|,|\eet| < 0.1$ which are consistent with  Eq.~\ref{tinynsi} and also the NSI phases in the allowed range, $\phi_{e\mu},\phi_{e\tau} \in (-\pi,\pi)$.  In addition, we explore the collective effect of the dominant NSI parameters ($\eem, \eet$) affecting the particular channels $\nu_\mu \to \nu _e$ (and  $\bar\nu_\mu \to \bar\nu_e$) so that the impact can be understood in totality. 

All the plots presented in this paper are obtained by using General Long baseline Experiment Simulator (GLoBES) software~\cite{Huber:2004ka,Huber:2007ji} which numerically 
solves  the full three flavour neutrino propagation equations using the PREM~\cite{Dziewonski:1981xy} density profile of the Earth, and the latest values of the neutrino parameters as obtained from global
fits~\cite{GonzalezGarcia:2012sz,Capozzi:2013csa,Forero:2014bxa}.

%----------------------------------------------------------------------%
\section{Probability expression for the $\nu_\mu \to \nu_ e$ channel}
\label{sec:cpvprob}
We give relevant analytic expressions in order to understand and contrast the features obtained from the probability in the case of vacuum and matter (both SI as well as NSI)~\cite{Arafune:1997hd,
Freund:2001pn,Akhmedov:2004ny,Gandhi:2004bj,Kimura:2002wd,Kimura:2002hb,Kimura:2006jj,Asano:2011nj,Chatterjee:2014gxa}.

 CP violation signal in a neutrino oscillation experiment such as DUNE 
 is characterised via a comparison of probabilities in the channel $\nu_\mu \to \nu_e$
 with its CP conjugate channel $\bar\nu_\mu \to \bar\nu_e$. If CP were conserved,  
 $\pmue{\delta} = \pmueb{\delta}$. 
  In order to quantify effects due to CP violation, we use the following observable CP asymmetry 
\begin{eqnarray} 
\label{eq:asymm}
\acp{\delta}  &=& \dfrac{\pmue{\delta} - \pmueb{\delta}}{\pmue{\delta} + \pmueb{\delta}}
%~\nonumber\\
%&=&
= \dfrac{\delpcp{\delta}}{\sumpcp{\delta}}~.
\end{eqnarray}
 It is important to note that though $\acp{\delta}$ is a good measure of CP violation, it has an obvious limitation in the sense that it cannot  allow us to deduce the source of CP violation effects. In the context of long baseline experiments where matter can induce fake CP effects, a non-zero value of $\acp{\delta}$ does not unequivocally imply intrinsic CP violation arising due to the Dirac CP phase. To get over the problem of differentiating between the case of CP violation due to intrinsic CP phase and CP violation arising due to the matter effect, other observables have been introduced~\cite{Ohlsson:2013ip} which can prove useful not only to establish whether CP violation effects arise purely due to the Dirac type CP phase or a combination of the intrinsic and extrinsic CP phases but also to distinguish between the cases based on spectral differences. In the present work, we are interested in bringing out the contribution coming from NSI towards the CP violation signal measured in terms of $\acp{\delta}$.

Let us consider the $\nu_\mu \to \nu_ e$ transition for propagation in vacuum and matter described below.
\subsection{{Review of \pmue{\delta} in vacuum :}}
 In vacuum, the  oscillation probability  for the $\nu_\mu \to \nu_ e$ channel is given by
\bea
\pmue{\delta}
&=& 
4(c_{13}^2 s_{23}^2 s_{13}^2 + {\cal J}_{}\sin{r_\lambda \lambda L})
\,\sin^2{\lambda L \over 2} \nonumber \\ 
&+&  2( c_{12} c_{23} c_{13}^2 s_{12} s_{23} s_{13} \cos{\delta} - 
c_{13}^2 s_{12}^2 s_{23}^2 s_{13}^2 )
\sin{r_\lambda \lambda L} \sin{\lambda L}
\nonumber \\ 
&+& 
4(c_{12}^2 c_{23}^2 c_{13}^2 s_{12}^2 +
c_{13}^2 s_{12}^4 s_{23}^2 s_{13}^2 -
2 c_{12} c_{23} c_{13}^2 s_{12}^3 s_{23} s_{13} \cos{\delta}
-
{\cal J}_{}\sin{\lambda L}) {\sin^2 {r_\lambda \lambda L \over 2}} 
\nonumber \\ 
&+& 8 ( c_{12} c_{23} c_{13}^2 s_{12} s_{23} s_{13} \cos{\delta} -
c_{13}^2 s_{12}^2 s_{23}^2 s_{13}^2 ) \sin^2{r_\lambda \lambda L \over 2} 
\sin^2{\lambda L \over 2} 
\label{eqmutoevac}
\eea
\noindent where ${\cal J}_{}= c_{12} c_{23} c_{13}^2 s_{12} s_{23}
s_{13}\sin {\delta}$  is an invariant 
that quantifies CP violation in the leptonic sector and is referred to as the Jarlskog invariant.  
The abbreviations $s_{ij}= \sin {{\theta}_{ij}}$, $c_{ij}=\cos {{\theta}_{ij}}$ are used in Eq.~\ref{eqmutoevac}.
For the CP-transformed channel (${\bar \nu}_\mu \to {\bar \nu}_e$), we need to replace $\delta_{} \to -\delta_{}$ in Eq.~\ref{eqmutoevac} to obtain $\pmueb{\delta}$.

The maximal $1-3$ mixing condition 
\begin{eqnarray}
\dfrac{L}{E} &=&  (2n-1) \dfrac{\pi}{2} ~ \frac{1}{1.267 \times \delta m^2_{31} (eV^2)} ~,
\end{eqnarray}
can be used to obtain the position of peaks in $\pmue{\delta}$ for $L=1300$ km (relevant for DUNE). $n=1,2,3,\ldots$ leads to $E^{\rm{peak}} \sim 2.5, 0.8, 0.5 \ldots~GeV$ for the first few peaks in vacuum probability. 

\subsection{{\pmue{\delta} in matter in presence of non-standard interactions :}}

The approximate expression for oscillation probability for $\numu \to \nue$ for NSI case can be obtained by retaining terms of  ${\cal O} (\epsilon_{\alpha\beta} s_{13})$, ${\cal O} (\epsilon_{\alpha\beta} r_\lambda)$ , ${\cal O} (s_{13} r_\lambda)$,  ${\cal O} (r_\lambda^2)$ and neglecting the higher order terms, 
\begin{eqnarray}
\pmue{\delta} &\simeq& 
 4 s_{13}^2 s_{23}^2 \left[\dfrac{ \sin^2 {(1-r_A )\lambda L/2}{}}{ (1-r_A)^2} \right]  \nonumber\\
 &+&  
 8  s_{13} s_{23} c_{23}
( |\epsilon_{e \mu}| c_{23} c_{\chi} - |\epsilon_{e \tau}| s_{23} c_{\omega} )  r_A \left[ \cos \frac{ \lambda L}{2} \dfrac{\sin {r_A \lambda L/2}}{r_A} 
  \dfrac{ \sin {(1 - r_A) \lambda L/2}}{(1-r_A)} 
 \right] \nonumber\\
  &+&
  8  s_{13} s_{23} c_{23}
( |\epsilon_{e \mu}| c_{23} s_{\chi} - |\epsilon_{e \tau}| s_{23} s_{\omega} ) r_A \left[\,\sin \frac{ \lambda L}{2} \dfrac{\sin {r_A \lambda L/2}}{r_A} 
  \dfrac{\sin {(1 - r_A) \lambda L/2}}{(1-r_A) } 
 \right]\nonumber\\
 &+&   8  s_{13} s_{23}^2 
( |\epsilon_{e \mu}| s_{23} c_{\chi} + |\epsilon_{e \tau}| c_{23} c_{\omega} ) r_A \left[
 \dfrac{\sin^2 {(1 - r_A) \lambda L/2} }{(1-r_A)^2} 
 \right] \ \nonumber\\
 &+&   4
 |\epsilon_{e\mu}| r_\lambda s_{2\times 12} c_{23}^3 \cos \phi_{e\mu}  r_A   \dfrac{\sin^2 r_A \lambda L/2 }{r_A^2} \nonumber\\
 &+&  2 
 |\epsilon_{e\mu}| r_\lambda s_{2\times 12} s_{23}^2 c_{23} \cos \phi_{e\mu} 
 r_A 
 \left[\cos \dfrac{ \lambda L}{2}\, \dfrac{\sin r_A \lambda L/2 }  {r_A} 
    \dfrac{\sin (1-r_A) \lambda L/2}{(1-r_A)} \right]
\nonumber\\
 &+&  
  |\epsilon_{e\mu}|  r_\lambda  s_{2\times 12} s_{23}^2 c_{23} \sin \phi_{e\mu} 
 r_A 
 \left[\sin \dfrac{ \lambda L}{2} \dfrac{\sin r_A \lambda L/2 }{r_A} 
    \dfrac{\sin (1-r_A) \lambda L/2}{(1-r_A)}  \right]
 \nonumber\\
  &-&   4 
 |\epsilon_{e\tau}|  r_\lambda  s_{2\times 12} c_{23}^2 {{s_{23} }}  \cos \phi_{e\tau} {r_A} \dfrac{\sin^2 r_A \lambda L/2}{r_A^2} \nonumber\\
 &+&   2
 |\epsilon_{e\tau}| r_\lambda  s_{2\times 12} s_{23}^2 c_{23} \cos \phi_{e\tau} r_A 
 \left[ \cos \dfrac{ \lambda L}{2} \dfrac{\sin r_A \lambda L/2 }{r_A} 
    \dfrac{\sin (1-r_A) \lambda L/2}{(1-r_A)} \right]
     \nonumber\\
 &+&  
  |\epsilon_{e\tau}| r_\lambda  s_{2\times 12} s_{23}^2 c_{23}  \sin \phi_{e\tau} r_A 
  \left[\sin \dfrac{\lambda L}{2} \dfrac{\sin r_A \lambda L/2 }{r_A}
    \dfrac{\sin (1-r_A) \lambda L/2}{(1-r_A)} \right]
  \nonumber\\
 &+& 8  r_\lambda  {\cal J}_r  \cos\delta 
 \left[\,\cos \dfrac{ \lambda L}{2}  \dfrac{\sin r_A \lambda L/2 }{r_A}
    \dfrac{\sin (1-r_A) \lambda L/2}{(1-r_A)} \right] \nonumber\\
  &-& 8 r_\lambda {\cal J}_r \sin \delta  \left[ \sin \dfrac{ \lambda L }{2} \sin \dfrac{ r_A \lambda L/2}{r_A} \dfrac{\sin (1-r_A) \lambda L/2}{(1-r_A)}  \right] \nonumber\\
 &+&  r_\lambda^2  c_{23}^2 s_{2 \times 12}^2   \dfrac{ \sin^2 (r_A \lambda L/2)}{r_A^2} \,,
 \label{pem}
 \end{eqnarray}
where ${\cal J}_{r} = {\cal J}/\sin\delta$ and $\chi=\phi_{e\mu}+ \delta$, $\omega = \phi_{e\tau}+\delta$.
  Note that only two parameters, $\epsilon_{e \mu} $ and $ \epsilon_{e\tau}$ enter in this 
leading order expression which implies that the rest of the NSI parameters are expected to play a sub-dominant role. 
The approximate expression (Eq.~\ref{pem}) allows us to illustrate the  qualitative impact of the moduli and phases of NSI terms  which  can in principle override effects due to the vacuum oscillation phase $\delta$ for certain choice of energies. 
 Also, the above expression is strictly valid when $r_\lambda \lambda L/2 \ll 1$  i.e. $L$ and $E$ are far away from the region where lower frequency oscillations dominate which is satisfied for long baseline experiments. 
For the case of DUNE, we have 
\begin{equation}
 r_\lambda \lambda L/2 = 0.125 \,\left[ 1.267 \times \dfrac{\delta m^2_{21}}{7.6 \times 10^{-5} ~eV^2} \dfrac{L}{1300~km} \dfrac{1	~GeV}{E} \right] < 1
 \label{lowfreq}
\end{equation}
Note that in addition to the vacuum oscillation frequency $\lambda L/2$  
\begin{equation}
\lambda L/2 = 4.0 \, \left[1.267 \times \dfrac{\delta m^2_{31}}{2.4 \times 10^{-3} ~eV^2} \dfrac{L}{1300~km} \dfrac{1	~GeV}{E} \right] 
\end{equation}
which has $E^{-1}$-dependence on energy,  matter (SI and NSI) introduces phase shifts such as $r_A  \lambda L/2$ (using $A = 0.756 \times 10^{-4} ~eV^2~ \rho~ (g/cc) ~E ~(GeV)$)
\begin{equation}
r_A \lambda L/2 = 0.4 \, \left[ 1.267 \times {0.756 \times 10^{-4}} \, \dfrac{\rho}{3.0 ~g/cc}\, \dfrac{L}{1300~km}\right] ,
\end{equation} 
which is $E$-independent and  hence achromatic.  The probability remains finite due to the $(1-r_A)$ and
$(1-r_A)^2$ terms in the denominator of Eq.~(\ref{pem}). 
Substituting $\delta \to -\delta$ and $r_A \to -r_A$ in $\pmue{\delta}$ (Eq.~\ref{pem}), we obtain $\pmueb{\delta}$. 
Let us discuss different limiting cases :

\begin{itemize}
\item {\sl{Vacuum ($r_A \to 0$) :}} \\
 When $r_A \to 0$, we recover the vacuum limit as expected~\cite{Akhmedov:2004ny}.
  In deriving Eq.~\ref{pem}, we have ignored these phase terms as $r_\lambda \lambda L/2 \ll 1$ (Eq.~\ref{lowfreq})  and hence we do not obtain vacuum  CP sensitivity as a limiting case Eq.~\ref{pem}. The correct expression to use in case of vacuum is given by Eq.~\ref{eqmutoevac} and it is evident that CP sensitivity arises due to higher order terms in $r_\lambda \lambda L/2$ in case of vacuum.  
 
\item {\sl{SI ($\eps_{\alpha\beta} \to 0$, $r_A \neq 0 $ and $r_\lambda \neq 0$) :}}\\ Only the first and last three lines of Eq.~\ref{pem} are non-zero in this case when only SI are operating~\cite{Kimura:2002wd,Kimura:2002hb}. The CP violation sensitivity is due to the terms proportional to $r_\lambda s_{13} \sim 0.03$ in this case as expected from standard matter case. When $r_A \to 1$, we are close to the resonance condition ($r_A=\cos 2\theta_{13}$ since $\theta_{13}$ is small).
\item {\sl{NSI-dominated regime ($r_\lambda \to 0$, $r_A \neq 0$ and $\epsilon_{e\mu},\epsilon_{e\tau} \neq 0$) : }}\\
 If we neglect terms of  ${\cal O} ( r_\lambda s_{13})$,  ${\cal O} (r_\lambda^2)$, we get the first four lines with non-zero terms in Eq.~\ref{pem}. The sole sensitivity to CP violating phase comes from the NSI terms. 
In this case, the CP violating effects appearing in second and third line are proportional to $s_{13} |\epsilon_{e\mu}|$ or $s_{13} |\epsilon_{e\tau}|$.  We can note that NSI effects with $|\epsilon_{\alpha\beta}| \simeq 0.1$ can in principle  override the standard CP violating effects in matter by one order of magnitude as $r_\lambda \simeq 0.03$. 

Another interesting aspect of this limit is that 
it would correspond to setting $\delta m^2_{21} =0$, which means we are effectively describing a two flavour case. In such a situation, as argued in Kikuchi et al.~\cite{Kikuchi:2008vq}, phase reduction is possible as one ends up with effectively one combination of phases such as $\chi=\delta+\phi_{e\mu}$ (or, $\omega=\delta + \phi_{e\tau}$) rather than individual phases $\delta$ and $\phi_{e\mu}$. This implies that if we are in the NSI dominated regime, it appears from the probability expression (Eq.~\ref{pem}) that there are degeneracies arising due to interplay of vacuum and NSI phases. We will refer to this as the {\sl{vacuum-NSI CP phase degeneracy}}. However, it turns out that once we take subdominant terms due to $r_\lambda \neq 0$ into consideration, the individual vacuum CP phase and NSI CP phase dependencies start showing up (through terms in lines 5-12 on RHS of Eq.~\ref{pem}) and such terms clearly aid in breaking of these vacuum-NSI CP phase degeneracies. 
\end{itemize}
It is worth mentioning that the chosen baseline of DUNE is sensitive to CP violation effects due to the intrinsic CP phase~\cite{Bass:2013vcg} as well as to additional CP violating effects arising due to SI or NSI. 

\begin{figure}[htb]
\centering
\includegraphics[width=\textwidth]
{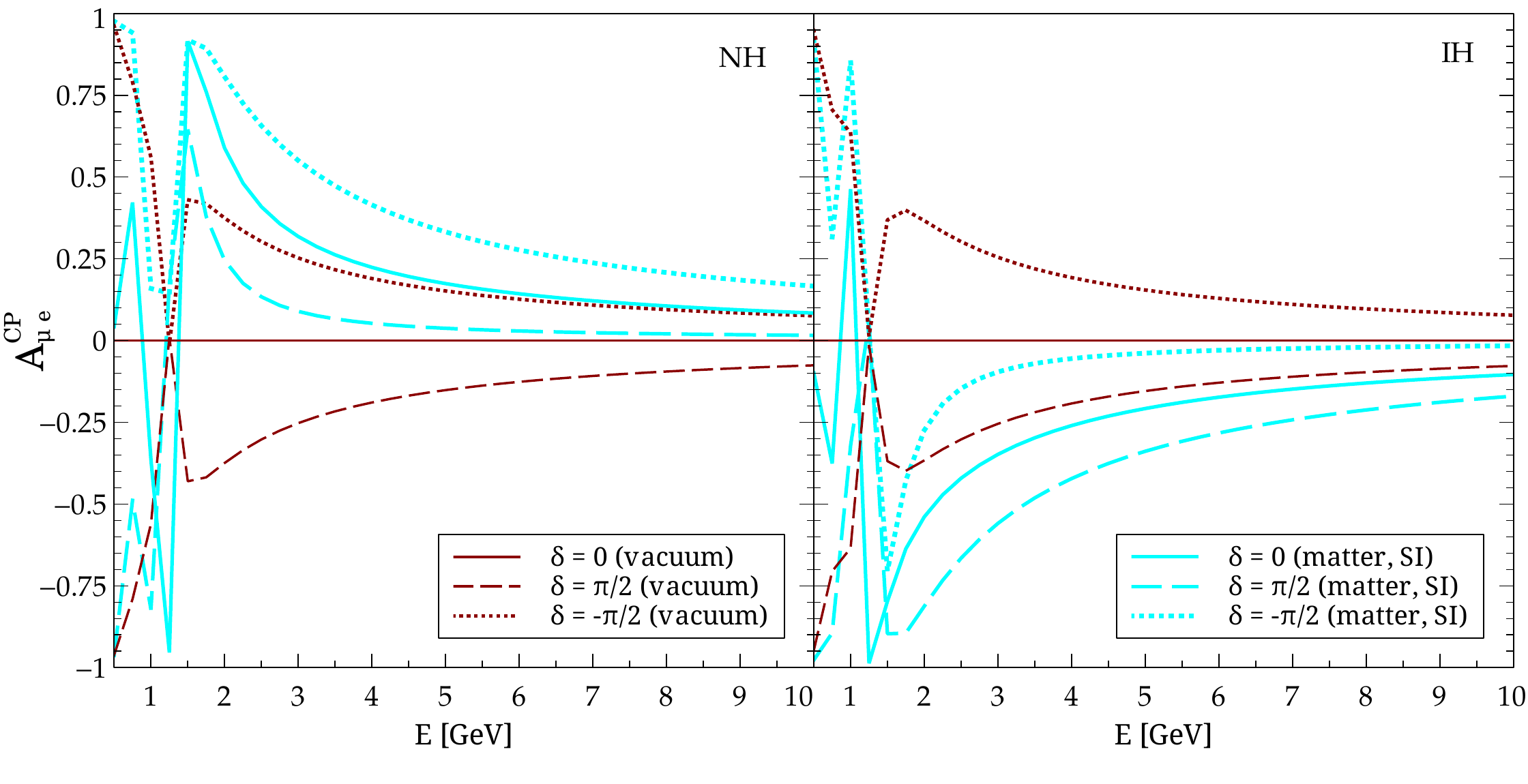}
\caption{\footnotesize{
Comparison of vacuum and  matter (SI) asymmetry, $\acp{\delta}$ for $L=1300$ km. The vacuum (matter, SI) case is shown in brown (cyan) for three different values of $\delta$ and for NH as well as IH. The solid, dashed and dotted lines correspond to $\delta=0$, $\delta= \pi/2$ and $\delta=-\pi/2$ respectively. 
 }}
\label{fig:acp_vac_si}
\end{figure}

\begin{figure}[h]
\centering
\includegraphics[width=0.85\textwidth]{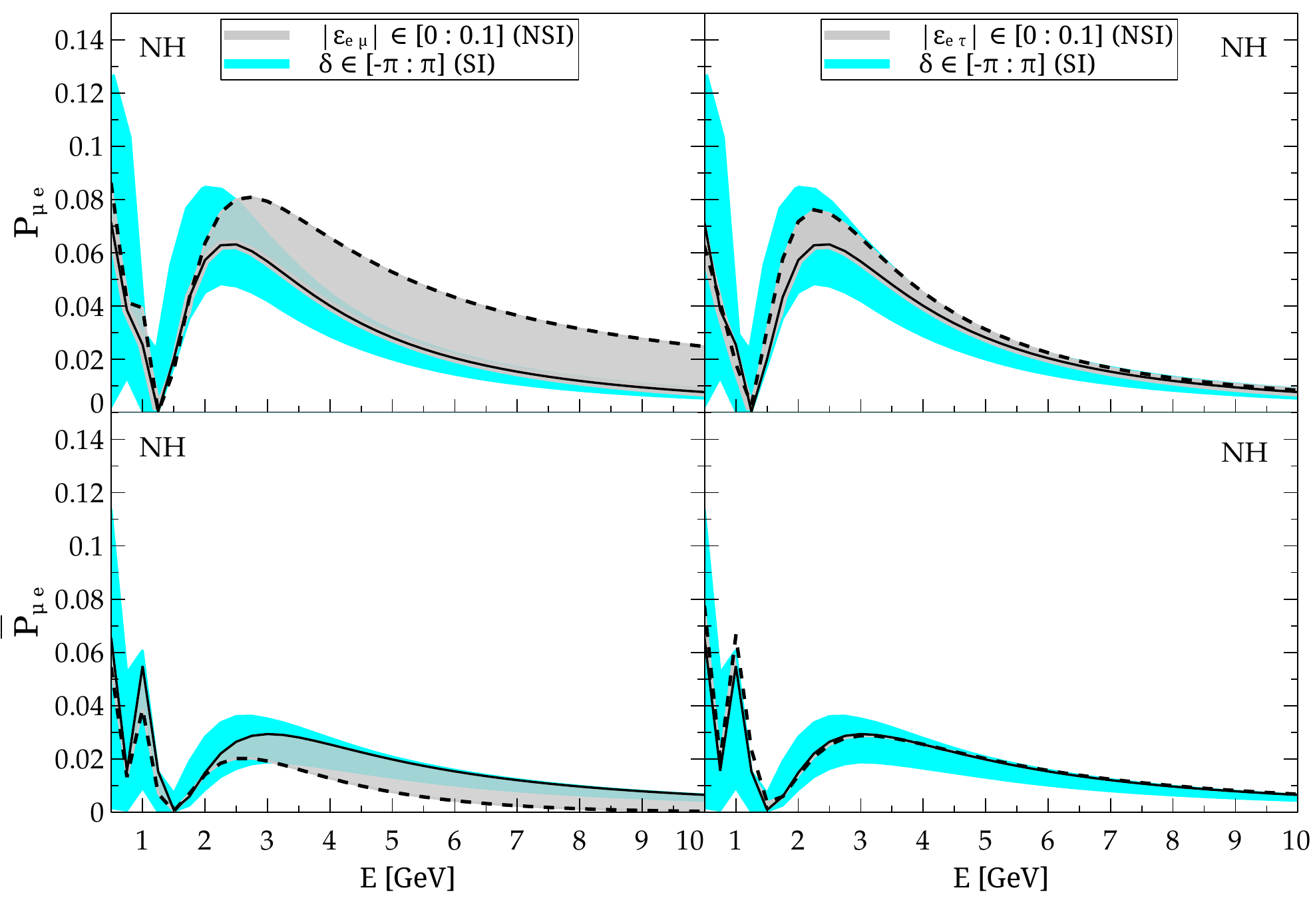}
\caption{\footnotesize{\pmue{\delta} and \pmueb{\delta} plotted as a function of energy for $L=1300$ km and the role of individual NSI parameters is depicted by varying the moduli of NSI parameters and assuming NH. 
The relevant moduli ($|\eem|$ on the left side or $|\eet|$ on the right side)  are varied in the allowed range (with all phases set to zero) as specified in the figure.  The cyan band corresponds to SI with $\delta \in (-\pi,\pi)$. The solid black line depicts the case of SI with $\delta=0$ and dashed black line depicts the case of NSI with $|\eem|=0.1$ in the left panel and $|\eet|=0.1$ in the right panel.  
 }}
\label{fig:prob_ind_amp}
\end{figure}

\begin{figure}[h]
\centering
\includegraphics[width=0.85\textwidth]{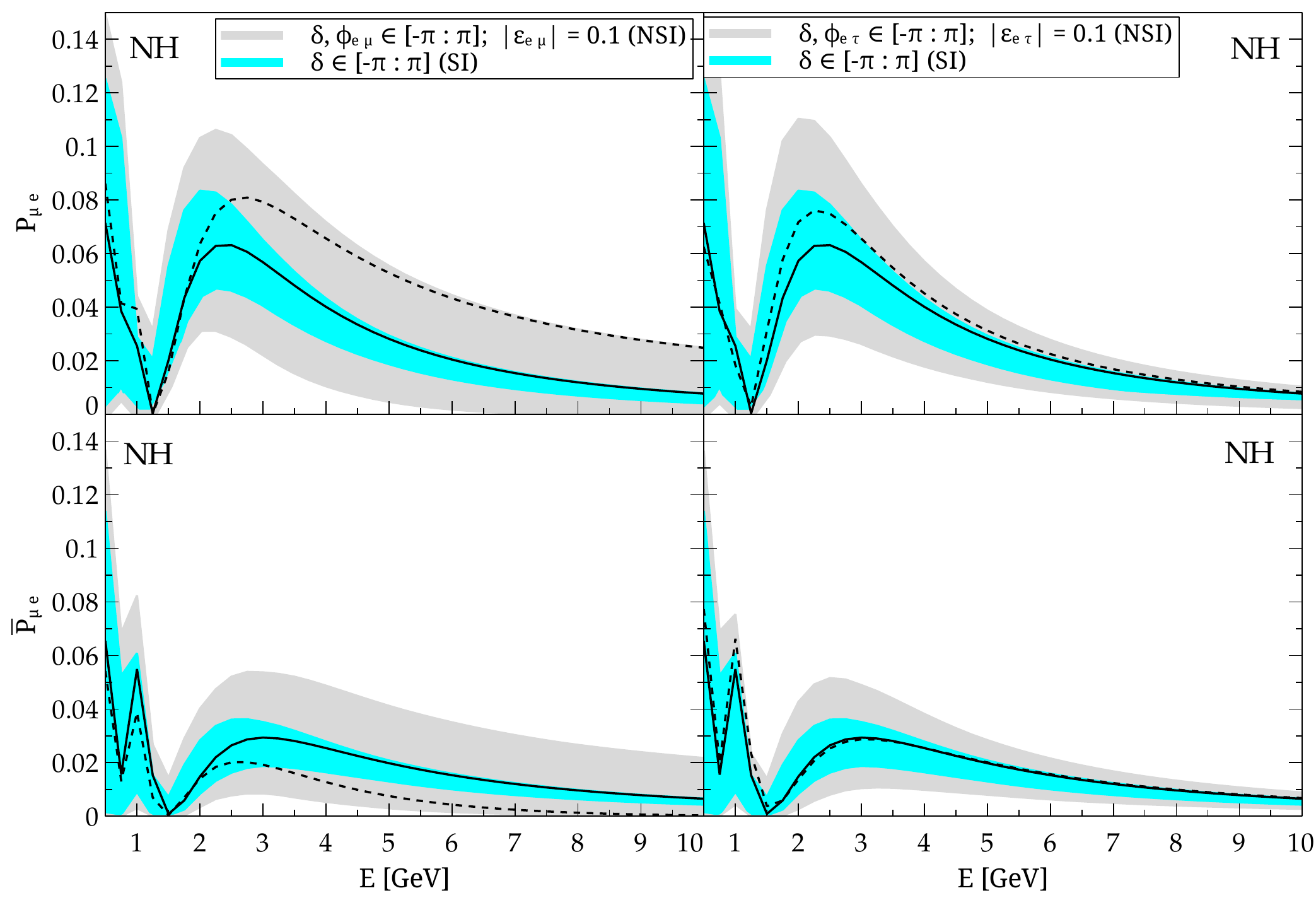}
\caption{\footnotesize{\pmue{\delta} and \pmueb{\delta} plotted as a function of energy for $L=1300$ km and the role of individual  NSI parameters is depicted by varying the phases of NSI parameters keeping the moduli fixed and assuming NH.  The relevant phases ($\delta,\phi_{e\mu}$ on the left side or $\delta,\phi_{e\tau}$ on the right side)  are varied in the allowed range as specified in the figure.  The cyan band corresponds to SI with $\delta \in (-\pi,\pi)$. The solid black line depicts the case of SI with $\delta=0$ and dashed black line depicts the case of NSI with $|\eem|=0.1$ in the left panel and $|\eet|=0.1$ in the right panel. 
 }}
\label{fig:prob_ind_phase}
\end{figure}

\begin{figure}[h]
\centering
\includegraphics[width=0.85\textwidth]{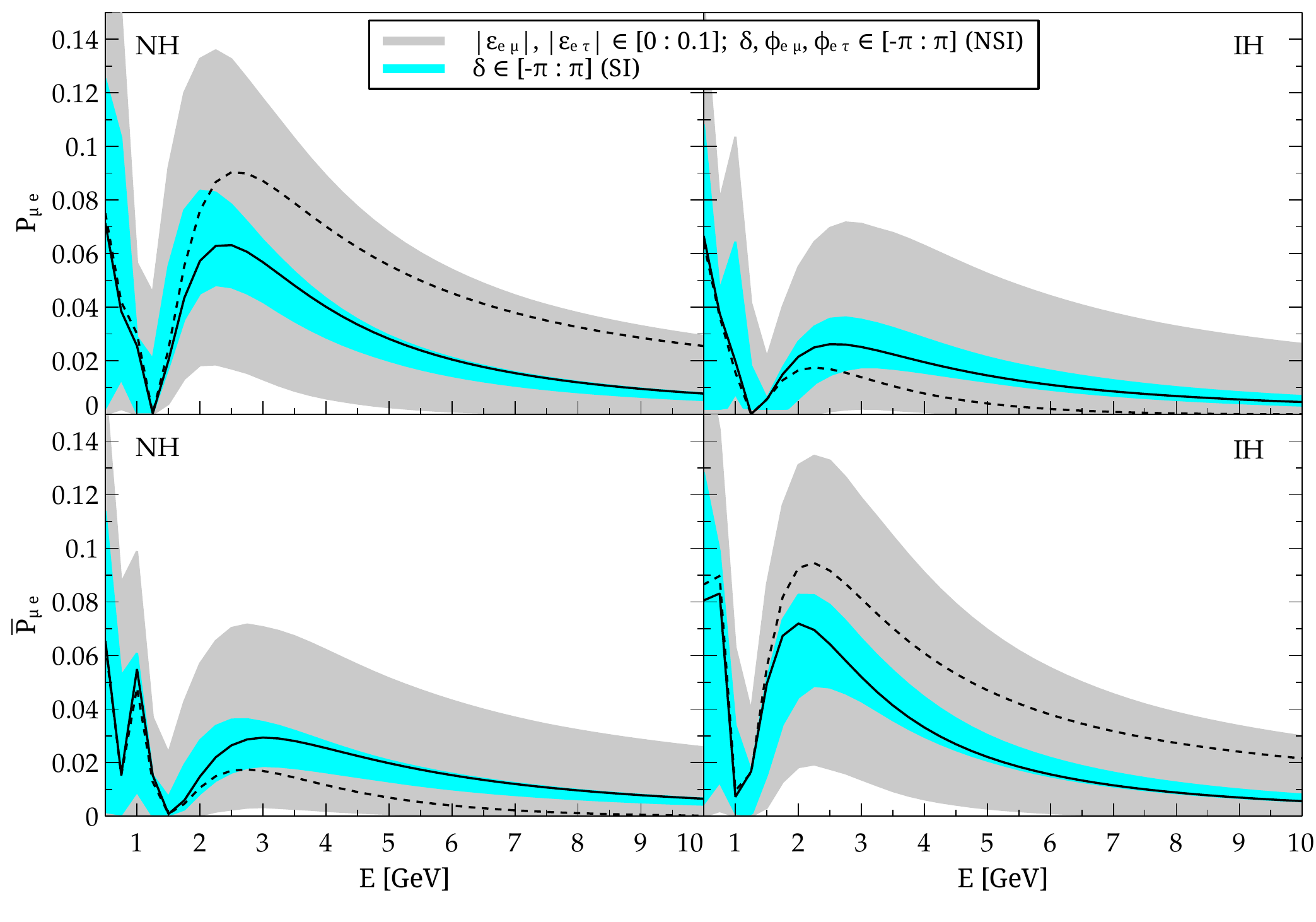}
\caption{\footnotesize{\pmue{\delta} and \pmueb{\delta} plotted as a function of energy for $L=1300$ km by considering the NSI parameters collectively and for NH (left panel) and IH (right panel). The phases ($\delta,\phi_{e\mu},\phi_{e\tau}$)  and  moduli  ($|\eem|,|\eet|$) are varied in the allowed range as specified in the figure.  The cyan band corresponds to SI with $\delta \in (-\pi,\pi)$. The solid black line depicts the case of SI with $\delta=0$ and dashed black line depicts the case of NSI with $|\eem|=|\eet|=0.1$.  
 }}
\label{fig:prob_collective}
\end{figure}

\begin{figure}[h]
\centering
\includegraphics[width=\textwidth]
{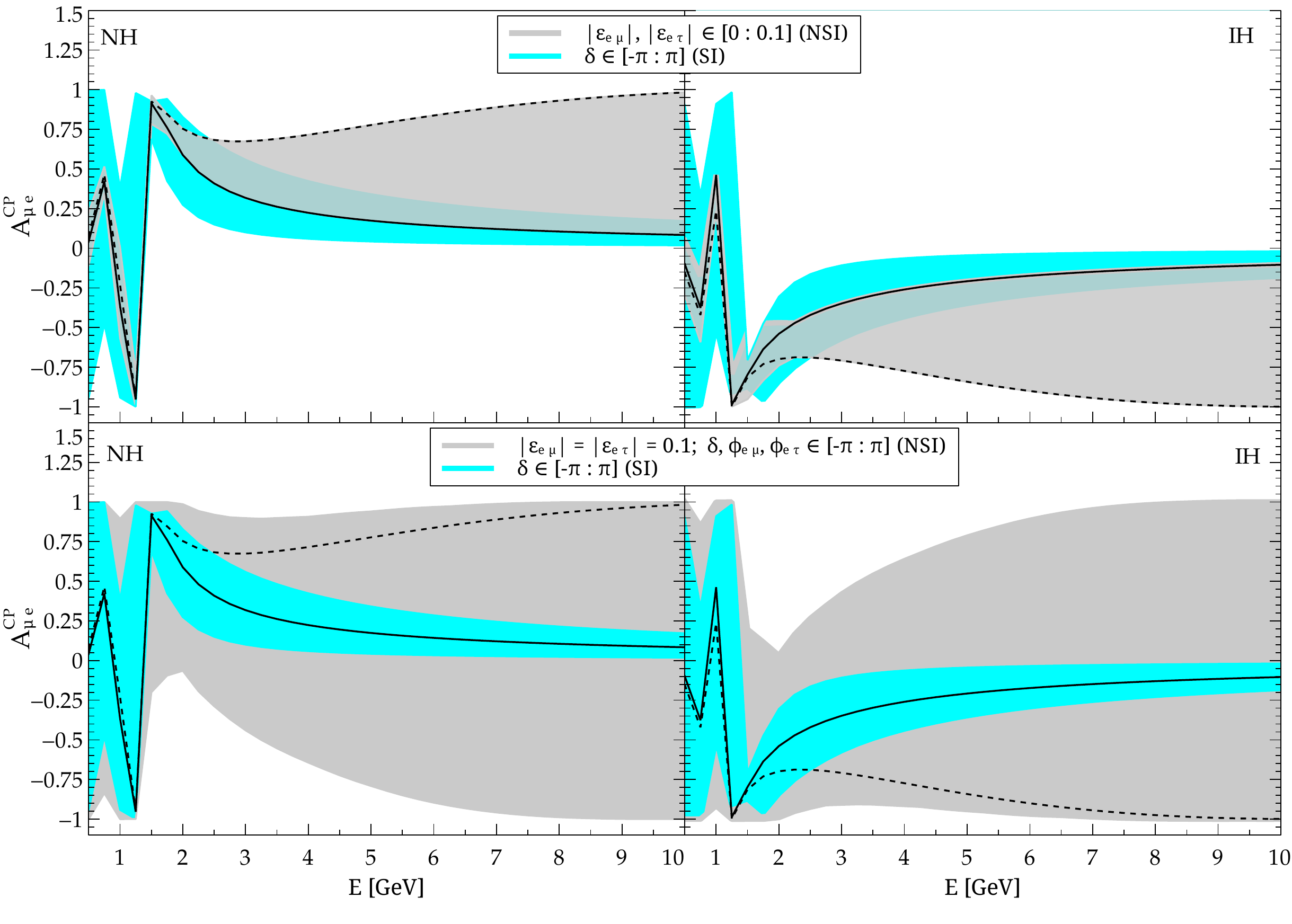}
\caption{\footnotesize{Impact of  collective NSI terms on the CP asymmetry bands as a function of energy for $L=1300$ km for NH and IH. 
Only the moduli of NSI parameters ($|\eem|,|\eet|$) are varied   in the top row and only 
 the phases ($\delta$,$\phi_{e\mu},\phi_{e\tau}$) for $|\eem|=|\eet|=0.1$  are varied in the allowed range as specified in the figure  in the bottom row.
  The cyan band corresponds to SI with $\delta \in (-\pi,\pi)$. The solid black line depicts the case of SI with $\delta=0$ and dashed black line depicts the case of NSI with $|\eem|=|\eet|=0.1$. 
  }}
\label{fig:acp_nsi_si_delfix}
\end{figure}

\section{Results and Discussion}
\label{sec:results}

Let us first discuss the case of vacuum. Using the CP-odd terms in Eq.~\ref{eqmutoevac}, the numerator in the  CP asymmetry (defined in Eq.~\ref{eq:asymm}) $\delpcp{\delta}$ is given by~\footnote{The denominator $\sumpcp{\delta}$ has the effect of rescaling the asymmetry curves.} 
\begin{eqnarray} 
\label{eq:a1_cp}
\delpcp{\delta}  &=&  8 \, {\cal J} \,\left[
 \sin (r_\lambda \lambda L) \sin^2 \frac{\lambda L}{2} - \sin (\lambda L) \sin^2 \frac{r_\lambda \lambda L}{2} \right] \nonumber \\ 
&=&   4 \, \sin\delta \, {\cal J}_r\, \left[ \sin\lambda L/2 \sin r_\lambda \lambda L/2 \sin {(1-r_\lambda) \lambda L/2} \right]~,
\end{eqnarray}
where the second line is obtained after rearranging the terms in the first line.  In order to have observable effects, we should have sizeable interference terms that involve the CP violating phase $\delta$. This implies that both $\lambda L/2$ as well as  $r_\lambda \lambda L/2$ must be taken into account. Naturally,  the $\acp{\delta}$ vanishes as $\delta \to 0,\pi$ and when $\delta = \pm \pi/2$,  $\acp{\delta}$ attains maximal values.
Also it can be noted that the normalised $\acp{\delta}$ grows linearly with $L/E$.

Using  Eq.~\ref{pem}, the numerator in the CP asymmetry (Eq.~\ref{eq:asymm}) for the SI ($\epsilon_{\alpha\beta} \to 0$ limit) can be expressed in a compact form 
\begin{eqnarray} 
\label{eq:a2_cp}
\delpcp{\delta}  &=&   8 \, r_\lambda {\cal J}_{} \, \dfrac{\sin r_A \lambda L/2}{r_ A}  \left[\Theta_- \cot \delta \cos \lambda L/2 + \Theta_+   \sin \lambda L/2 \right] ~,
 \end{eqnarray}
 where 
 $\Theta_{\pm} = \sin [(r_A-1) \lambda L/2]/(r_A-1) \pm \sin [(r_A+1)\lambda L/2]/(r_A+1)$. The CP sensitivity comes from terms proportional to $r_\lambda$ in this case. In contrast to the vacuum expression, the $\acp{0} \neq 0$ and  this can be attributed to the fake CP effects arising due to matter being CP asymmetric.  In the limit $r_A \to 0$, one would expect non-zero vacuum terms to remain (see Eq.~\ref{eq:a1_cp}) but we get $\acp{\delta}\to 0$ if we use Eq.~\ref{eq:a2_cp}. 
 This is due to the fact that we have neglected phase terms proportional to $r_\lambda \lambda L/2$ (in view of Eq.~\ref{lowfreq}) while deriving Eq.~\ref{pem} and Eq.~\ref{eq:a2_cp}.

A comparison of $\acp{\delta}$ in vacuum and in matter (SI) is shown for fixed values of $\delta_{CP} =0, \pm \pi/2$  in Fig.~\ref{fig:acp_vac_si}. The features of the curves obtained in case of vacuum for  $\delta = 0$ and $\delta=\pm \pi/2$ can be understood from Eq.~\ref{eq:a1_cp}. 
For non-zero value of $\delta$, the $\acp{\delta}$ in vacuum is expected to grow as $L/E$ which implies for a fixed $L$, the $\acp{\delta}$ is large for small values of $E$. 
 It is seen that $\delta=\pi/2$ (dashed brown line) leads to $\acp{\delta} < 0$  and $\delta=-\pi/2$ (dotted  brown line)  leads to $\acp{\delta} > 0$, independent of the hierarchy, which follows from Eq.~\ref{eq:a1_cp}.
However, in case of SI, hierarchy dependent effects are visible. Even when $\delta =0$, we notice that SI introduces a non-zero asymmetry (solid cyan line) as opposed to vacuum expectation (solid brown line). In case of NH, the $\acp{\delta} > 0$ for $E > 1.5 ~GeV$ while for IH, the $\acp{\delta} < 0$ for $E > 1.5 ~GeV$. Also, in the high energy region, $\delta = -\pi/2$ ($\delta=\pi/2$) leads to large asymmetry in case of NH (IH) in comparison to the SI case. 
We also show two cases of non-zero $\delta$ ($=\pm \pi/2$). For NH, $\delta=\pi/2$ diminishes the $\acp{\delta}$ value (dashed cyan line) while $\delta = -\pi/2$ increases the value of $\acp{\delta}$ (dotted cyan line) in comparison to $\delta=0$ case (solid cyan line). Similar trend is seen in case of IH. The effects in case of SI can be explained using Eq.~\ref{eq:a2_cp}.
The fake CP asymmetry should be substantial when $L/E$ is large such that matter effects are important. This leads to large asymmetry in the low energy region for a fixed baseline. 
Also, near the peak position of the probability (where the flux is large), we note that  large asymmetry is introduced due to SI.
 The asymmetry in case of SI shows a saturation behaviour on positive side  ($\acp{\delta} >0$) at energies beyond $4~GeV$ for NH while for IH a similar saturation behaviour is seen for energies beyond $4~GeV$ but on the negative side ($\acp{\delta} < 0$) for $\delta = \pm \pi/2$.  
This comes about due to the presence of $E$-independence of phase term $r_A \lambda L/2$ in Eq.~\ref{eq:a2_cp}. 

For the general case of NSI,  writing analytic expression for asymmetry is cumbersome  with all terms present in Eq.~\ref{pem}. We discuss the differences arising due to the two leading non-zero NSI parameters, viz., $\eem,\eet$ affecting $\nu_\mu \to \nu_e$ transition by a visual 
comparison of the plots for NSI case with those for the SI case (see Figs.~\ref{fig:prob_ind_amp},\ref{fig:prob_ind_phase} and \ref{fig:prob_collective}). As $\eem,\eet$ are complex, we discuss visible effects arising due to the variation in 
absolute value and phase of these terms. In order to understand all the subtle effects arising due to non-zero NSI terms, we first discuss the isolated case (one NSI parameter non-zero at a time) and then  collective case (the relevant NSI parameters are taken non-zero simultaneously). 
\begin{enumerate}
\item {\sl{Impact of individual NSI terms :}} \\ Here only one NSI parameter is taken to be non-zero at a time.  Fig.~\ref{fig:prob_ind_amp} and \ref{fig:prob_ind_phase} depict the effect due to individual NSI terms. The cyan band represents the SI case ($\delta \in (-\pi:\pi)$) while the grey band is for NSI.  
Any non-zero value of the NSI parameters leads to a deviation from the SI band (cyan band). As expected, both the oscillation amplitude as well as the position of peaks/dips change in presence of NSI (grey band) in comparison to the SI case with $\delta \in (-\pi:\pi)$ (cyan band). 
The solid line (black) shows the SI case ($\delta = 0$) while the dashed line (black) depicts the case of NSI with all phases set to zero for either $|\eem|=0.1$ or $|\eet|=0.1$.

 The effect of non-zero moduli of NSI terms, $|\eem|$ and $|\eet|$ is shown in Fig.~\ref{fig:prob_ind_amp}.   The grey bands correspond to $|\eem| \in (0-0.1)$ or $|\eet| \in (0-0.1)$. The lower limit of $0$ corresponds to standard matter effects while the upper limit of $0.1$ is consistent with Eq.~\ref{tinynsi}. 
 The effect of $|\eem|  $ ($|\eet|$) is shown on the left (right) hand side of Fig.~\ref{fig:prob_ind_amp}. $|\eem|$ has larger impact than $|\eet|$ on \pmue{\delta} and \pmueb{\delta} in case of NH (see  Fig.~\ref{fig:prob_ind_amp}).  It can be seen from the top row in Fig.~\ref{fig:prob_ind_amp} that as long as the $|\eem| >0 $ or $|\eet| > 0$, the $\pmue{\delta}$ is in general larger than the solid black line (SI, $\delta=0$) for neutrinos and NH. For the same range, we can note that $\pmueb{\delta}$ is lower than the  black solid line  for anti-neutrinos and NH (bottom row in  Fig.~\ref{fig:prob_ind_amp}). 
This shift in peak position is clearly visible in case of $|\eem|$, the peak of neutrino (anti-neutrino) channel in case of NSI is shifted to the right (left) side of the SI peak for NH.  For IH, the neutrinos and antineutrinos interchange roles  and the peak of neutrino (anti-neutrino) channel in case of NSI is shifted to the left (right) side of the SI peak. 
Note that for NH, $|\eet|$ has smaller effect on the $\pmue{\delta}$ and $\pmueb{\delta}$ in comparison to $|\eem|$. 
This is because this parameter comes with a negative sign in Eq.~\ref{pem} (second and third line) and therefore 
 large changes are expected due to this parameter mainly for IH and $\pmueb{\delta}$.

The effect of varying the relevant phases in the allowed range ($\delta \in (-\pi:\pi)$, $\phi_{e\mu} \in (-\pi:\pi)$, $\phi_{e\tau} \in (-\pi:\pi)$) is shown in Fig.~\ref{fig:prob_ind_phase}. If we take $|\eem| \neq 0$, we will have two phases $\delta,\phi_{e\mu}$ (left hand side of Fig.~\ref{fig:prob_ind_phase}) and if we take $|\eet| \neq 0$, we will have two phases $\delta, \phi_{e\tau}$ (right hand side of Fig.~\ref{fig:prob_ind_phase}). For the case of NSI, we have used $|\eem|=0.1$ and $|\eet| = 0.1$ in the left and right panel respectively. The general effect of varying the phases is that it leads to a band (cyan) around the solid black line in case of SI. For NSI case, there is further broadening of  bands (shown in grey) on both sides of the SI (cyan) band. From Eq.~\ref{pem}, the terms in lines 2-10 are responsible for the grey band.

\item {\sl{Collective impact of NSI terms :}}
The collective case where both the NSI parameters are taken non-zero simultaneously ($|\eem|=0-0.1$ , $|\eet|=0-0.1$ and $\delta \in (-\pi:\pi)$, $\phi_{e\mu} \in (-\pi:\pi)$, $\phi_{e\tau} \in (-\pi:\pi)$
) is shown in Fig.~\ref{fig:prob_collective}. 
 It turns out the the combined effect of the variation of phases and moduli of two NSI parameters acts in the same direction and overall we get larger visible differences at the level of probability in comparison to the SI case due to a non-trivial interplay of moduli and phase parts of the two NSI terms. So, considering the dominant NSI parameters simultaneously is crucial as it allows for better distinguishability of the NSI from SI. 
For IH (right hand side of Fig.~\ref{fig:prob_collective}), the sign of NSI terms will be reversed  and we expect opposite effects in comparison to NH. 
\end{enumerate}

\begin{figure}[h]
\centering
\includegraphics[width=\textwidth]
{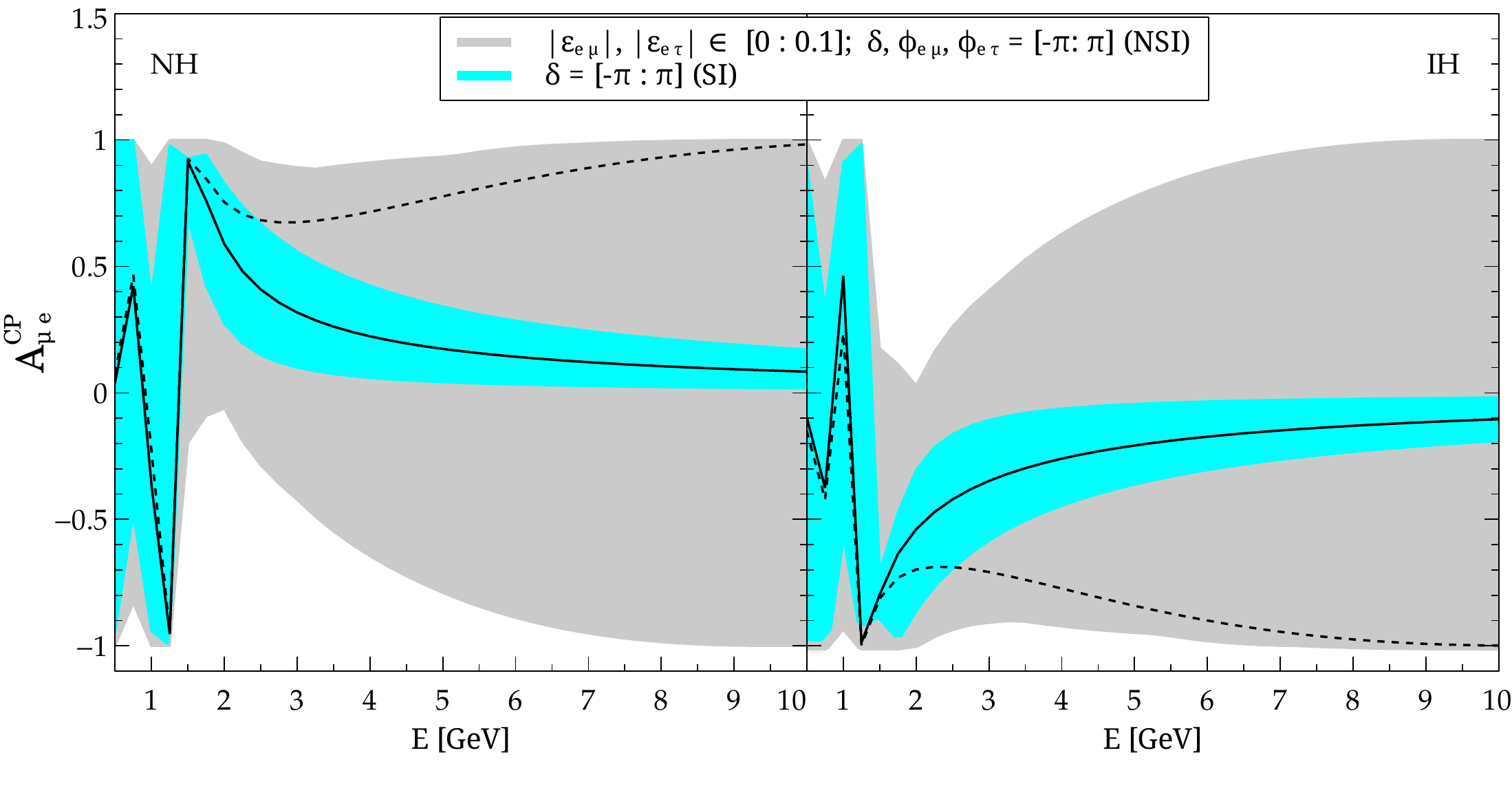}
\caption{\footnotesize{
Impact of  collective NSI terms on the CP asymmetry bands as a function of energy for $L=1300$ km for NH (left) and IH (right). 
Both moduli  ($|\eem|,|\eet|$) and 
  phases ($\delta$,$\phi_{e\mu},\phi_{e\tau}$) are varied in the allowed range as specified in the figure (see also Fig.~\ref{fig:prob_collective}). 
  The cyan band corresponds to SI with $\delta \in (-\pi,\pi)$. The solid black line depicts the case of SI with $\delta=0$ and dashed black line depicts the case of NSI with $|\eem|=|\eet|=0.1$. %
}}
\label{fig:acp_dcp_non0_da_non0}
\end{figure}

Having discussed the imprint of individual and collective  NSI terms on the $\pmue{\delta}$ and $\pmueb{\delta}$, let us now turn to the CP violating observable, i.e. the $\acp{\delta}$. In Fig.~\ref{fig:acp_nsi_si_delfix} and Fig.~\ref{fig:acp_dcp_non0_da_non0},  the maximum and minimum $\acp{\delta}$ is plotted and the SI (NSI) cases are  shown as cyan (grey) bands as a function of $E$ for the two hierarchies. The maximum and minimum values  are obtained by varying the relevant vacuum (the Dirac phase) and  NSI parameters (moduli and/or phases) in the allowed range mentioned in the figure caption.
 Once again the cyan band corresponds to the case of SI and the grey band corresponds to NSI. The solid black line depicts the SI curve for $\delta=0$ and dashed black line depicts the collective NSI curve  ($|\eem|=0.1$ and $|\eet|=0.1$). 
 The discussion of $\pmue{\delta}$ and $\pmueb{\delta}$ above is reflected in the $\acp{\delta}$ curves. 
The top row in Fig.~\ref{fig:acp_nsi_si_delfix} shows the impact of changing the modulus of collective NSI terms while the bottom row shows the impact of changing the phase of the  NSI terms keeping the modulus fixed  ($|\eem|=0.1$ and $|\eet|=0.1$). The grey band in the top row depicts the fake CP effect in the same spirit as effects arising due to SI while the grey band in the bottom row contain fake CP effect {\it{a~la}} SI along with spurious contributions to the intrinsic CP phase arising due to non-zero NSI phases.  
For $E < 1.5 ~GeV$, the two curves corresponding to SI and NSI almost coincide. The difference in cyan and grey bands increases with increasing the energy beyond $1.5~ GeV$. 
 The most dramatic  aspect of  NSI is that the naive argument of needing large $L/E$ (and hence small $E$ for fixed value of $L$) in order to obtain large asymmetries due to matter effects does not work any more and we can obtain large asymmetries even at large energies, as is reflected from the asymmetry plots (Fig.~\ref{fig:acp_nsi_si_delfix} and \ref{fig:acp_dcp_non0_da_non0}). The case of NH and IH show opposite sign of asymmetry in the grey band (top row). However, when phases are taken into account, the asymmetry may also change sign at higher energies (bottom row).

Finally, Fig.~\ref{fig:acp_dcp_non0_da_non0} shows the maximum and minimum asymmetry for the case where both moduli and phases of NSI parameters $\eem,\eet$ are varied simultaneously. {\sl{Large values of \acp{\delta} at higher energies vindicates the presence of non-zero NSI terms.}}

\section{Event rates and CP violation sensitivity at DUNE far detector}
\label{sec:events}

Our simulations are carried out using the General Long baseline Experiment Simulator (GLoBES) software~\cite{Huber:2004ka,Huber:2007ji} using the best-fit values of the standard oscillation parameters~\cite{GonzalezGarcia:2012sz,Capozzi:2013csa,Forero:2014bxa}. We assume a particular hierarchy (NH or IH) and allow the value of $\delta \in [-\pi,\pi]$. We describe the reference set up as follows. 
The beam power  is 1.2 MW and a wide band neutrino beam is obtained in energy range, $E=0.5-10$ GeV.  The far detector comprises of an (unmagnetised) LArTPC which has excellent energy % as well as angular
 resolution
 both for muons and electrons. The specifications for the
Liquid Argon TPC detector are given in Table~\ref{tab:exptparams}. We shall
assume here five year operation period, or, equivalently, an effective
exposure of $175 (=35 \times 5 )$  kt-yr in the neutrino mode and  175 kt-yr separately in the anti-neutrino mode.

%-------------

\begin{table}[h]
\centering
\begin{tabular}{ l  l}
\hline
&\\
Beam power & 1.2 MW \\
Fiducial detector mass & 35 kt\\
Bin width & 250 MeV \\
\multirow{2}{*}{Exposure} 
  & $35 \times 5 =$ 175 kt-yr for $\nu$\\
  &  $35 \times 5 =$ 175 kt-yr for $\bar{\nu}$ \\
Energy Resolution ($\sigma (E)$) & $\sqrt{(0.01)^{2} + (0.15)^{2}/(yE_{}) + (0.03)^{2}}$\\

Detector efficiency ($\mathcal{E}$) & $85\%$ \\

&\\
\hline
\end{tabular}
\caption{\label{tab:exptparams} The LArTPC far-detector specifications used in our simulations for DUNE.}
\end{table}

\begin{figure}[h]
\centering
\includegraphics[width=0.85\textwidth]{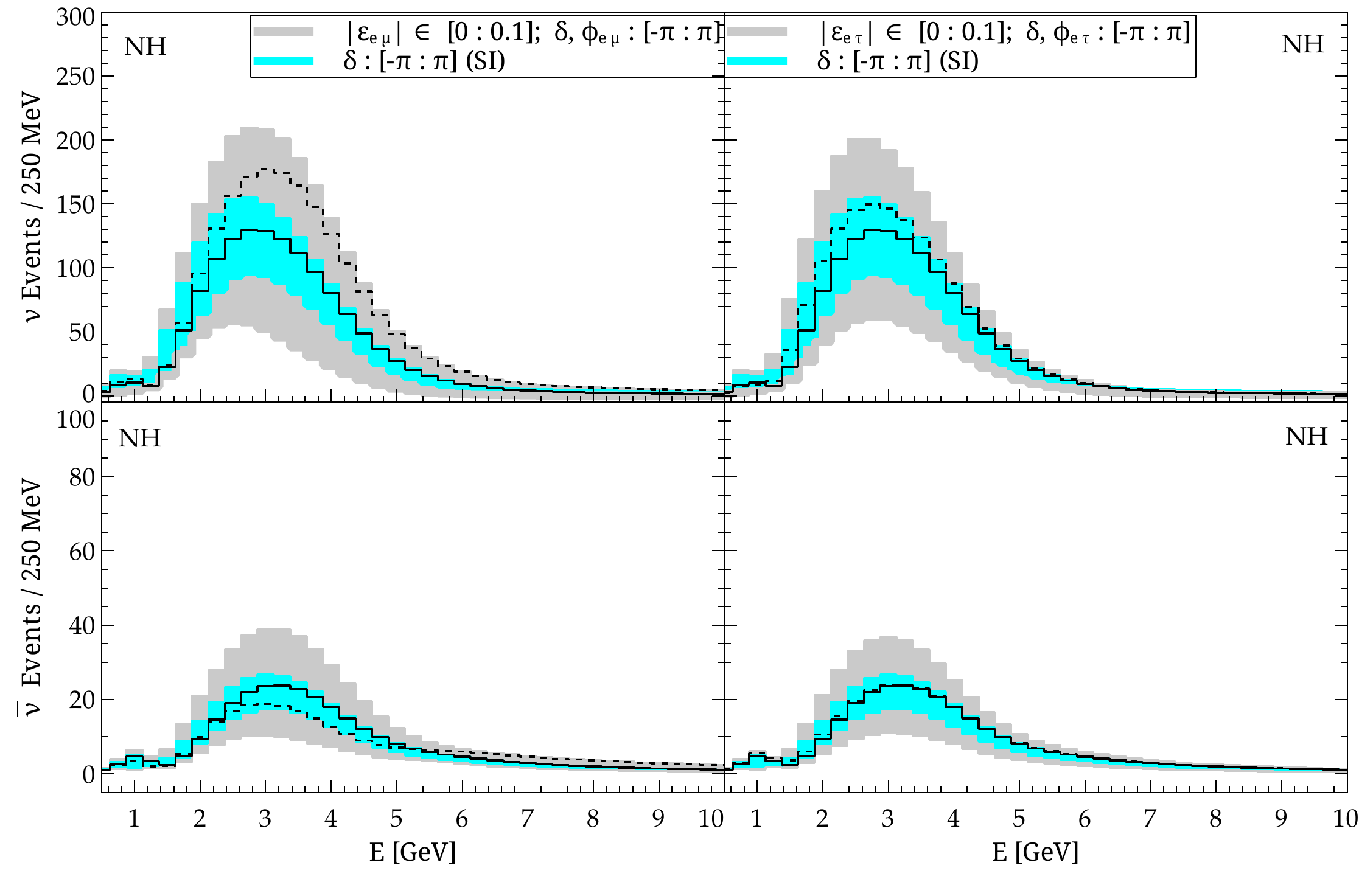}
\caption{\footnotesize{Impact of individual NSI  terms on the $\nu$ and $\bar\nu$ events plotted as a function of energy at DUNE far detector for NH (see also Fig.~2 and 3). The cyan band depicts the case of SI with $\delta \in [-\pi : \pi]$ while the grey band depicts the case of NSI with variation in the moduli and phase of the relevant NSI parameter as specified in the figure. The black solid (dashed) line depicts the case of  SI with $\delta=0$  (NSI with all phases set to zero).
 }}
\label{fig:nu_anu_event_spread_em_et}
\end{figure}

\begin{figure}[h]
\centering
\includegraphics[width=0.85\textwidth]{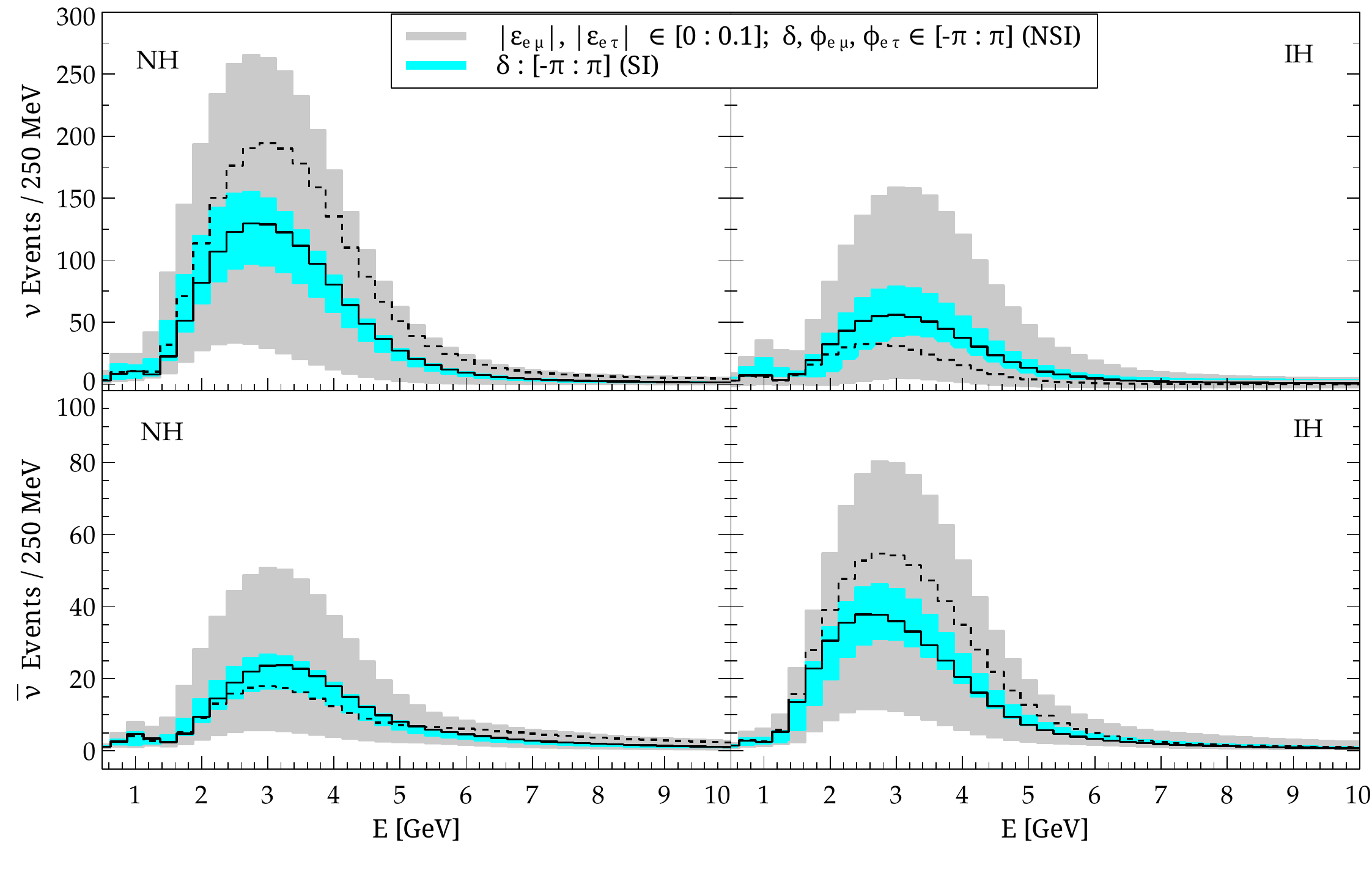}
\caption{\footnotesize{Impact of collective NSI  terms on the $\nu$ and $\bar\nu$ events plotted as a function of energy at DUNE far detector for NH and IH (see also Fig.~4 and 6). The cyan band depicts the case of SI with $\delta \in (-\pi : \pi)$ while the grey band depicts the case of NSI with variation in the moduli and phase of the two NSI parameters as specified in the figure. The black solid (dashed) line depicts the case of  SI with $\delta=0$  (NSI with all phases set to zero).}}
\label{fig:nu_anu_event_spread_em_et2}
\end{figure}

\begin{figure}[htb]
\centering
\includegraphics[width=0.85\textwidth]{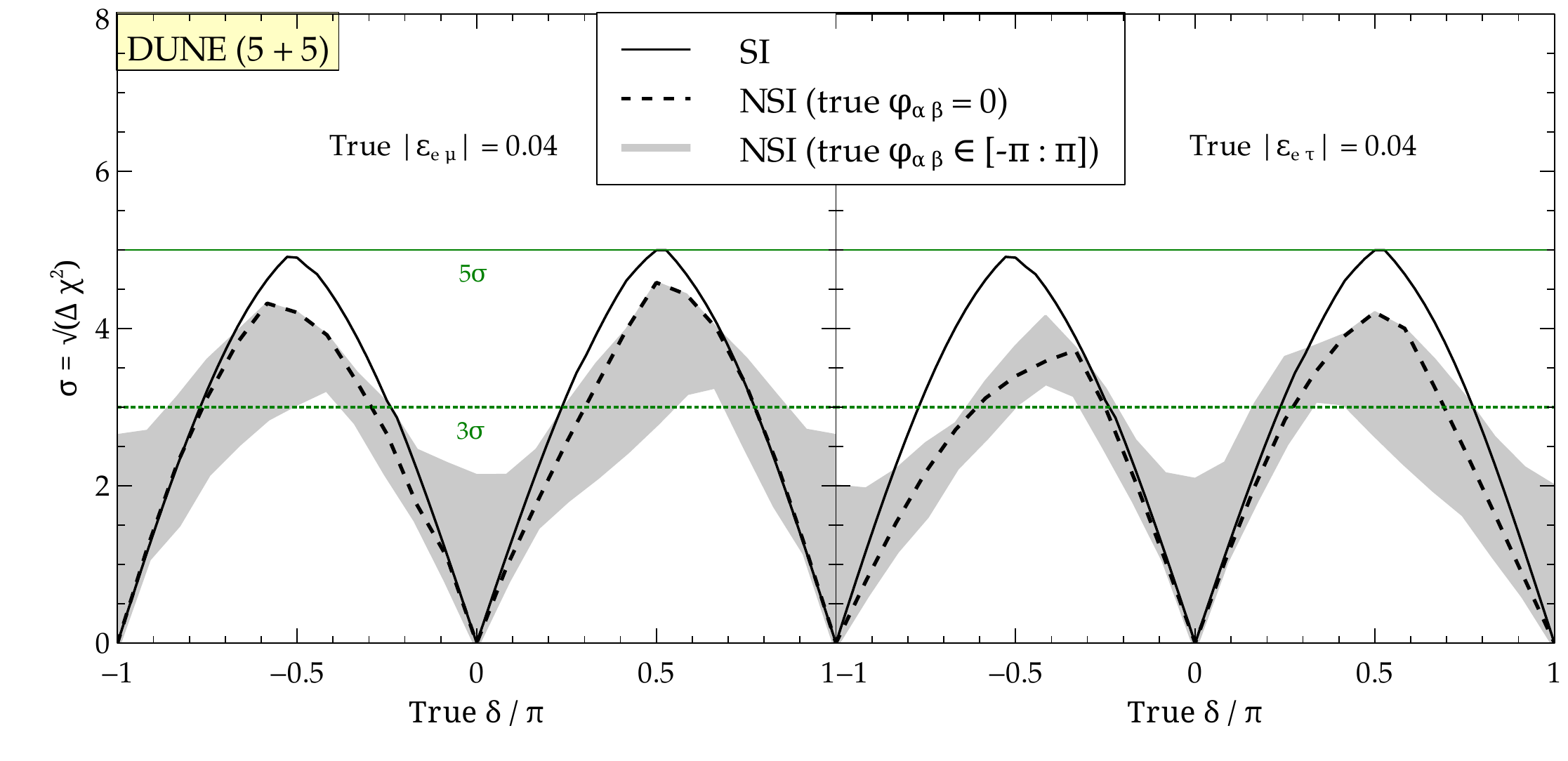}
\caption{\footnotesize{CP violation sensitivity plotted as a function of  true $\delta$ by using the appearance  ($\nu_\mu \to \nu_e$) channel. The impact of 
$\varepsilon_{e\mu}$ ($\varepsilon_{e\tau}$) is shown in the left (right) plot. The grey band depicts the case of  NSI with variation in the moduli and phase of the relevant NSI parameter as specified in the figure. The black solid (dashed) line depicts the case of  SI   (NSI with true values of all the NSI phases set to zero). 
}}
\label{fig:newfig}
\end{figure}

In Fig.\ \ref{fig:nu_anu_event_spread_em_et}, we show electron neutrino and antineutrino event histograms for SI and NSI for different values of $\eema$ and $\eeta$ for the reference set-up mentioned above. 
These histograms show the maximum spread in the events when the relevant parameters (in case of SI, $\delta$ and in case of NSI, moduli and phases of NSI parameters) are varied over the allowed range specified in the figure. The region between maximum and the minimum event rates is shown as shaded cyan (grey) band in case of SI (NSI).  It should be noted that the set of varied oscillation parameters (SI phase or NSI phases) that give the spread in the event rates in each bin in Fig.~\ref{fig:nu_anu_event_spread_em_et} are in general different. In Fig~\ref{fig:nu_anu_event_spread_em_et2}, we show the impact of NSI terms collectively and with variation both in moduli and phases. As an example, we show how the individual  off-diagonal NSI terms ($\varepsilon_{e\mu}$ and $\varepsilon_{e\tau}$) impact the CP violation sensitivity at DUNE in Fig.~\ref{fig:newfig}.
We observe the following
\begin{itemize}
\item The event spectrum follows the probability plots and there is large difference near the first oscillation maximum of $\nu_\mu \to \nu_e$ probability even in the standard matter case which gets enhanced in presence of NSI. 
\item Though the asymmetries are fairly large in the higher energy regime, the number of events are not so large in high energy bins due to the small flux at those energies. This holds for NH as well as IH. 
\item
There are overlapping regions where SI and NSI results are consistent with one another due to a certain favourable choice of parameters. We refer to these as the NSI-SI degeneracies. Due to presence of these new degeneracies, it becomes hard to ascribe the signal to SI alone or to NSI. The two bands corresponding to  SI and NSI phase variations respectively are overlapping for a wide range of energies including those where the flux peaks.
\item Neutrino and anti-neutrino event spectrum  looks very similar and this due to the fact that we have taken  phase variations into account which tend to hide some of the cancellation effects coming about if we take only the moduli of relevant NSI terms (for instance, the lower right panel of Fig. 4 shows one such cancellation effect). 
\item If we take the relevant NSI parameters collectively and vary both moduli and phases, the effect adds up constructively for NH and IH (see Fig~\ref{fig:nu_anu_event_spread_em_et2}). 
\item We show the impact of individual NSI terms on the CP violation sensitivity at DUNE in Fig.~\ref{fig:newfig} for a nominal exposure of $350$ kt.MW.yr and runtime of five  years each in neutrino and antineutrino mode. The sensitivity to CP violation signals how well we can 
  demarkate the CP violating values from the CP conserving ones ($0,\pm\pi$). 
  The $3\sigma$ and $5\sigma$ values are shown as horizontal lines.   It can be seen that presence of NSI can impact the sensitivity to CP violation more prominently near the peaks or the dips.  The most drastic effect is that NSI can mimic CP violation even for the CP conserving values ($0,\pm\pi$).    
    For a 
  detailed study on CP violation sensitivity at DUNE in presence of NSI, see Ref.~\cite{Masud:2016bvp}.
\end{itemize}

\section{Conclusion}
\label{sec:conclude}

We have studied the role played by NSI on the issue of determining the leptonic CP violation in the context of DUNE. We use the  probability expression for $\nu_\mu \to \nu_e$ channel in presence of NSI (by retaining the contribution of  terms such as $|\epsilon_{\alpha\beta}| s_{13},|\eet| r_\lambda,  r_\lambda s_{13}, r_\lambda^2$ which appear at similar order in Eq.~\ref{pem}). It turns out that only two NSI parameters appear in the approximate expression and retaining all the terms is crucial in order to understand the breaking of the vacuum-NSI phase degeneracies. 
Since the two NSI parameters ($|\eet|,|\eem|$) are complex, we consider (fixed and variable) absolute value of the NSI terms consistent with present constraints as well as vary their phases to obtain insight into the role played by these quantities. These additional moduli and phases can lead to large changes in the asymmetries at DUNE. Significant differences are seen in the probability and asymmetry curves near the peak of the $\pmue{\delta}$ and $\pmueb{\delta}$  and at higher energies in comparison to the SI case for a certain choice of parameters. In terms of asymmetries, we depict the role of moduli and phases separately as well as in conjunction when individual (collective) NSI terms are taken into account. 
Again large changes are visible in asymmetry plots when we exploit the full variation in parameters. In case of ordinary matter effects one expects that the asymmetry should be large for large $L/E$ which implies small $E$ for fixed $L$ as it would allow for matter effects to develop. But when NSI is switched on, one sees large changes also in higher energy bins due to the interplay of phases and moduli of these terms as well as with the CP phase $\delta$. 
 However in terms of event rates, there is a certain wash out of effects seen especially in the higher energy range.
This is because the flux falls off and hence the effects appearing in higher energy range are not cleanly observable. 
From the event rate analysis, we show that the alteration in event rates is most significant near the peak but exists all across  the spectrum. There is no significant distortion in the spectrum shape. 

Interestingly, there are overlapping regions where SI and NSI results are consistent with one another due to a certain favourable choice of parameters. We refer to these as the NSI-SI degeneracies. Due to presence of these new degeneracies, it becomes hard to ascribe the signal to SI alone or to NSI. The two bands corresponding to  SI and NSI phase variations respectively are overlapping for a wide range of energies including those where the flux peaks (in case of very small $|\epsilon_{\alpha\beta}| \sim 0.01$, the degeneracy is perfect).

It is shown that any conclusion regarding the CP phase can not be arrived at without a 
thorough analysis of correlations and degeneracies that arise due to the extra parameters 
in addition to the standard three flavour parameters where the only source of CP violation is 
the Dirac phase. Thus in order to ascribe any result on the CP phase to the lone CP phase in the standard three neutrino paradigm, it is crucial to rule out new physics scenarios that can contribute to the signal. Within NSI scenario itself we showed that there are two parameters with both magnitude and phase that can vary over wide range and can contribute to it. In addition there can be other scenarios, such as the recently discussed 
 sterile neutrinos~\cite{Berryman:2015nua,Gandhi:2015xza}  which also add new parameters into the framework and can mimic the same event spectrum. So, one needs to devise a mechanism by which one can ascertain that other scenarios can be ruled out in order to conclude anything about the CP phase to complete the standard framework. 

{\bf{Note added : }} After our work was submitted to the preprint arXiv, 
some papers~\cite{deGouvea:2015ndi,Coloma:2015kiu,Liao:2016hsa,Forero:2016cmb} appeared on the preprint arxiv in which the consequences of NSI were investigated in the context of DUNE. However none of them deal with the precise question that we have addressed in the present work. 

\section*{Acknowledgements} 
It is a pleasure to thank Raj Gandhi for useful discussions and critical comments on the manuscript. 
We acknowledge Sushant Raut for help related to NSI implementation into GLOBES and involvement of Suprabh Prakash during the initial stages of this work.  
 We acknowledge the use of HRI cluster facility to carry out computations in this work. AC acknowledges support from Jaehoon Yu, Amir Farbin and Intensity Frontier Programme of Fermilab.
PM acknowledges support from University Grants Commission under the second phase of University with Potential of Excellence at JNU and partial support from the European Union grant FP7 ITN INVISIBLES (Marie Curie Actions, PITN-GA-2011-289442). PM thanks the organisers of IPP15, Tehran for  warm hospitality  during the final stages of  this work.

\bibliographystyle{apsrev}
\bibliography{referencesnsi}

\begin{thebibliography}{10}
\expandafter\ifx\csname bibnamefont\endcsname\relax
  \def\bibnamefont#1{#1}\fi
\expandafter\ifx\csname bibfnamefont\endcsname\relax
  \def\bibfnamefont#1{#1}\fi
\expandafter\ifx\csname url\endcsname\relax
  \def\url#1{\texttt{#1}}\fi
\expandafter\ifx\csname urlprefix\endcsname\relax\def\urlprefix{URL }\fi
\providecommand{\bibinfo}[2]{#2}
\providecommand{\eprint}[2][]{\url{#2}}

\bibitem{nobel2013}
\bibinfo{author}{\bibfnamefont{F.}~\bibnamefont{Englert}} \bibnamefont{and}
  \bibinfo{author}{\bibfnamefont{P.~W.} \bibnamefont{Higgs}},
  \emph{\bibinfo{title}{For the theoretical discovery of a mechanism that
  contributes to our understanding of the origin of mass of subatomic
  particles, and which recently was confirmed through the discovery of the
  predicted fundamental particle, by the atlas and cms experiments at cern's
  large hadron collider}}, \bibinfo{note}{the Nobel Prize in Physics 2013.}

\bibitem{nobel2015}
\bibinfo{author}{\bibfnamefont{T.}~\bibnamefont{Kajita}} \bibnamefont{and}
  \bibinfo{author}{\bibfnamefont{A.~B.} \bibnamefont{McDonald}},
  \emph{\bibinfo{title}{For the discovery of neutrino oscillations, which shows
  that neutrinos have mass}}, \bibinfo{note}{the Nobel Prize in Physics 2015.}

\bibitem{GonzalezGarcia:2012sz}
\bibinfo{author}{\bibfnamefont{M.}~\bibnamefont{Gonzalez-Garcia}},
  \bibinfo{author}{\bibfnamefont{M.}~\bibnamefont{Maltoni}},
  \bibinfo{author}{\bibfnamefont{J.}~\bibnamefont{Salvado}}, \bibnamefont{and}
  \bibinfo{author}{\bibfnamefont{T.}~\bibnamefont{Schwetz}},
  \bibinfo{journal}{JHEP} \textbf{\bibinfo{volume}{1212}}, \bibinfo{pages}{123}
  (\bibinfo{year}{2012}), \eprint{1209.3023}.

\bibitem{Capozzi:2013csa}
\bibinfo{author}{\bibfnamefont{F.}~\bibnamefont{Capozzi}},
  \bibinfo{author}{\bibfnamefont{G.}~\bibnamefont{Fogli}},
  \bibinfo{author}{\bibfnamefont{E.}~\bibnamefont{Lisi}},
  \bibinfo{author}{\bibfnamefont{A.}~\bibnamefont{Marrone}},
  \bibinfo{author}{\bibfnamefont{D.}~\bibnamefont{Montanino}}, \emph{et~al.},
  \bibinfo{journal}{Phys.Rev.} \textbf{\bibinfo{volume}{D89}},
  \bibinfo{pages}{093018} (\bibinfo{year}{2014}), \eprint{1312.2878}.

\bibitem{Forero:2014bxa}
\bibinfo{author}{\bibfnamefont{D.~V.} \bibnamefont{Forero}},
  \bibinfo{author}{\bibfnamefont{M.}~\bibnamefont{Tortola}}, \bibnamefont{and}
  \bibinfo{author}{\bibfnamefont{J.~W.~F.} \bibnamefont{Valle}},
  \bibinfo{journal}{Phys. Rev.}
  \textbf{\bibinfo{volume}{D90}}(\bibinfo{number}{9}), \bibinfo{pages}{093006}
  (\bibinfo{year}{2014}), \eprint{1405.7540}.

\bibitem{juno}
\bibinfo{author}{\bibfnamefont{Y.-F.} \bibnamefont{Li}},
  \bibinfo{author}{\bibfnamefont{J.}~\bibnamefont{Cao}},
  \bibinfo{author}{\bibfnamefont{Y.}~\bibnamefont{Wang}}, \bibnamefont{and}
  \bibinfo{author}{\bibfnamefont{L.}~\bibnamefont{Zhan}},
  \bibinfo{journal}{Phys. Rev. D} \textbf{\bibinfo{volume}{88}},
  \bibinfo{pages}{013008} (\bibinfo{year}{2013}).

\bibitem{accelerator}
\bibinfo{author}{\bibfnamefont{T.}~\bibnamefont{Nakaya}} \bibnamefont{and}
  \bibinfo{author}{\bibfnamefont{R.~K.} \bibnamefont{Plunkett}},
  \bibinfo{journal}{Arxiv eprints}  (\bibinfo{year}{2015}),
  \eprint{1507.08134}.

\bibitem{nova}
\bibinfo{author}{\bibfnamefont{D.~S.} \bibnamefont{Ayres}} \emph{et~al.}
  (\bibinfo{collaboration}{NOvA Collaboration}), \bibinfo{journal}{Arxiv
  eprints}  (\bibinfo{year}{2004}), \eprint{hep-ex/0503053}.

\bibitem{2013arXiv1307.7335L}
\bibinfo{author}{\bibfnamefont{C.}~\bibnamefont{Adams}} \emph{et~al.}
  (\bibinfo{collaboration}{LBNE Collaboration}), \bibinfo{journal}{ArXiv
  e-prints}  (\bibinfo{year}{2013}), \eprint{1307.7335}.

\bibitem{Bass:2013vcg}
\bibinfo{author}{\bibfnamefont{M.}~\bibnamefont{Bass}} \emph{et~al.}
  (\bibinfo{collaboration}{LBNE Collaboration}), \bibinfo{journal}{Phys. Rev.}
  \textbf{\bibinfo{volume}{D91}}, \bibinfo{pages}{052015}
  (\bibinfo{year}{2015}), \eprint{1311.0212}.

\bibitem{icalreport}
\bibinfo{author}{\bibfnamefont{S.}~\bibnamefont{Ahmed}} \emph{et~al.}
  (\bibinfo{collaboration}{ICAL Collaboration}), \bibinfo{journal}{Arxiv
  eprints}  (\bibinfo{year}{2015}), \eprint{1505.07380}.

\bibitem{pingu}
\bibinfo{author}{\bibfnamefont{M.~G.} \bibnamefont{Aartsen}} \emph{et~al.}
  (\bibinfo{collaboration}{IceCube PINGU Collaboration}),
  \bibinfo{journal}{Arxiv eprints}  (\bibinfo{year}{2014}), \eprint{1401.2046}.

\bibitem{Arafune:1996bt}
\bibinfo{author}{\bibfnamefont{J.}~\bibnamefont{Arafune}} \bibnamefont{and}
  \bibinfo{author}{\bibfnamefont{J.}~\bibnamefont{Sato}},
  \bibinfo{journal}{Phys. Rev.} \textbf{\bibinfo{volume}{D55}},
  \bibinfo{pages}{1653} (\bibinfo{year}{1997}), \eprint{hep-ph/9607437}.

\bibitem{Tanimoto:1996ky}
\bibinfo{author}{\bibfnamefont{M.}~\bibnamefont{Tanimoto}},
  \bibinfo{journal}{Phys. Rev.} \textbf{\bibinfo{volume}{D55}},
  \bibinfo{pages}{322} (\bibinfo{year}{1997}), \eprint{hep-ph/9605413}.

\bibitem{Bilenky:1997dd}
\bibinfo{author}{\bibfnamefont{S.~M.} \bibnamefont{Bilenky}},
  \bibinfo{author}{\bibfnamefont{C.}~\bibnamefont{Giunti}}, \bibnamefont{and}
  \bibinfo{author}{\bibfnamefont{W.}~\bibnamefont{Grimus}},
  \bibinfo{journal}{Phys. Rev.} \textbf{\bibinfo{volume}{D58}},
  \bibinfo{pages}{033001} (\bibinfo{year}{1998}), \eprint{hep-ph/9712537}.

\bibitem{Barger:1980jm}
\bibinfo{author}{\bibfnamefont{V.~D.} \bibnamefont{Barger}},
  \bibinfo{author}{\bibfnamefont{K.}~\bibnamefont{Whisnant}}, \bibnamefont{and}
  \bibinfo{author}{\bibfnamefont{R.~J.~N.} \bibnamefont{Phillips}},
  \bibinfo{journal}{Phys. Rev. Lett.} \textbf{\bibinfo{volume}{45}},
  \bibinfo{pages}{2084} (\bibinfo{year}{1980}).

\bibitem{Diwan:2003bp}
\bibinfo{author}{\bibfnamefont{M.~V.} \bibnamefont{Diwan}} \emph{et~al.},
  \bibinfo{journal}{Phys. Rev.} \textbf{\bibinfo{volume}{D68}},
  \bibinfo{pages}{012002} (\bibinfo{year}{2003}), \eprint{hep-ph/0303081}.

\bibitem{Brahmachari:2003bk}
\bibinfo{author}{\bibfnamefont{B.}~\bibnamefont{Brahmachari}},
  \bibinfo{author}{\bibfnamefont{S.}~\bibnamefont{Choubey}}, \bibnamefont{and}
  \bibinfo{author}{\bibfnamefont{P.}~\bibnamefont{Roy}},
  \bibinfo{journal}{Nucl. Phys.} \textbf{\bibinfo{volume}{B671}},
  \bibinfo{pages}{483} (\bibinfo{year}{2003}), \eprint{hep-ph/0303078}.

\bibitem{Akhmedov:2004ve}
\bibinfo{author}{\bibfnamefont{E.~K.} \bibnamefont{Akhmedov}},
  \bibinfo{journal}{Phys. Scripta} \textbf{\bibinfo{volume}{T121}},
  \bibinfo{pages}{65} (\bibinfo{year}{2005}), \eprint{hep-ph/0412029}.

\bibitem{Marciano:2006uc}
\bibinfo{author}{\bibfnamefont{W.}~\bibnamefont{Marciano}} \bibnamefont{and}
  \bibinfo{author}{\bibfnamefont{Z.}~\bibnamefont{Parsa}},
  \bibinfo{journal}{Nucl. Phys. Proc. Suppl.} \textbf{\bibinfo{volume}{221}},
  \bibinfo{pages}{166} (\bibinfo{year}{2011}), \eprint{hep-ph/0610258}.

\bibitem{pakvasa}
\bibinfo{author}{\bibfnamefont{S.}~\bibnamefont{Pakvasa}},
  \bibinfo{journal}{Journal of Physics: Conference Series}
  \textbf{\bibinfo{volume}{556}}(\bibinfo{number}{1}), \bibinfo{pages}{012060}
  (\bibinfo{year}{2014}).

\bibitem{Davidson:2008bu}
\bibinfo{author}{\bibfnamefont{S.}~\bibnamefont{Davidson}},
  \bibinfo{author}{\bibfnamefont{E.}~\bibnamefont{Nardi}}, \bibnamefont{and}
  \bibinfo{author}{\bibfnamefont{Y.}~\bibnamefont{Nir}},
  \bibinfo{journal}{Phys. Rept.} \textbf{\bibinfo{volume}{466}},
  \bibinfo{pages}{105} (\bibinfo{year}{2008}), \eprint{0802.2962}.

\bibitem{cabibbo}
\bibinfo{author}{\bibfnamefont{N.}~\bibnamefont{Cabibbo}},
  \bibinfo{journal}{Phys. Rev. Lett.} \textbf{\bibinfo{volume}{10}},
  \bibinfo{pages}{531} (\bibinfo{year}{1963}).

\bibitem{km}
\bibinfo{author}{\bibfnamefont{M.}~\bibnamefont{Kobayashi}} \bibnamefont{and}
  \bibinfo{author}{\bibfnamefont{T.}~\bibnamefont{Maskawa}},
  \bibinfo{journal}{Progress of Theoretical Physics}
  \textbf{\bibinfo{volume}{49}}(\bibinfo{number}{2}), \bibinfo{pages}{652}
  (\bibinfo{year}{1973}).

\bibitem{Pakvasa:1975ti}
\bibinfo{author}{\bibfnamefont{S.}~\bibnamefont{Pakvasa}} \bibnamefont{and}
  \bibinfo{author}{\bibfnamefont{H.}~\bibnamefont{Sugawara}},
  \bibinfo{journal}{Phys. Rev.} \textbf{\bibinfo{volume}{D14}},
  \bibinfo{pages}{305} (\bibinfo{year}{1976}).

\bibitem{pdg2014}
\bibinfo{author}{\bibfnamefont{K.~A.} \bibnamefont{Olive}} \emph{et~al.}
  (\bibinfo{collaboration}{Particle Data Group}), \bibinfo{journal}{Chin.
  Phys.} \textbf{\bibinfo{volume}{C38}}, \bibinfo{pages}{090001}
  (\bibinfo{year}{2014}).

\bibitem{pontecorvo}
\bibinfo{author}{\bibfnamefont{B.}~\bibnamefont{Pontecorvo}},
  \bibinfo{journal}{Sov. Phys. JETP} \textbf{\bibinfo{volume}{26}},
  \bibinfo{pages}{984} (\bibinfo{year}{1968}).

\bibitem{mns}
\bibinfo{author}{\bibfnamefont{Z.}~\bibnamefont{Maki}},
  \bibinfo{author}{\bibfnamefont{M.}~\bibnamefont{Nakagawa}}, \bibnamefont{and}
  \bibinfo{author}{\bibfnamefont{S.}~\bibnamefont{Sakata}},
  \bibinfo{journal}{Progress of Theoretical Physics}
  \textbf{\bibinfo{volume}{28}}(\bibinfo{number}{5}), \bibinfo{pages}{870}
  (\bibinfo{year}{1962}).

\bibitem{Bilenky:1984fg}
\bibinfo{author}{\bibfnamefont{S.~M.} \bibnamefont{Bilenky}},
  \bibinfo{author}{\bibfnamefont{N.~P.} \bibnamefont{Nedelcheva}},
  \bibnamefont{and} \bibinfo{author}{\bibfnamefont{S.~T.}
  \bibnamefont{Petcov}}, \bibinfo{journal}{Nucl. Phys.}
  \textbf{\bibinfo{volume}{B247}}, \bibinfo{pages}{61} (\bibinfo{year}{1984}).

\bibitem{Wolfenstein:1977ue}
\bibinfo{author}{\bibfnamefont{L.}~\bibnamefont{Wolfenstein}},
  \bibinfo{journal}{Phys. Rev.} \textbf{\bibinfo{volume}{D17}},
  \bibinfo{pages}{2369} (\bibinfo{year}{1978}).

\bibitem{Mikheev:1987qk}
\bibinfo{author}{\bibfnamefont{S.~P.} \bibnamefont{Mikheev}} \bibnamefont{and}
  \bibinfo{author}{\bibfnamefont{A.~Y.} \bibnamefont{Smirnov}},
  \bibinfo{journal}{Sov. Phys. Usp.} \textbf{\bibinfo{volume}{30}},
  \bibinfo{pages}{759} (\bibinfo{year}{1987}).

\bibitem{Arafune:1997hd}
\bibinfo{author}{\bibfnamefont{J.}~\bibnamefont{Arafune}},
  \bibinfo{author}{\bibfnamefont{M.}~\bibnamefont{Koike}}, \bibnamefont{and}
  \bibinfo{author}{\bibfnamefont{J.}~\bibnamefont{Sato}},
  \bibinfo{journal}{Phys. Rev.} \textbf{\bibinfo{volume}{D56}},
  \bibinfo{pages}{3093} (\bibinfo{year}{1997}), \bibinfo{note}{[Erratum: Phys.
  Rev.D60,119905(1999)]}, \eprint{hep-ph/9703351}.

\bibitem{Berryman:2015nua}
\bibinfo{author}{\bibfnamefont{J.~M.} \bibnamefont{Berryman}},
  \bibinfo{author}{\bibfnamefont{A.}~\bibnamefont{de~Gouvea}},
  \bibinfo{author}{\bibfnamefont{K.~J.} \bibnamefont{Kelly}}, \bibnamefont{and}
  \bibinfo{author}{\bibfnamefont{A.}~\bibnamefont{Kobach}},
  \bibinfo{journal}{Arxiv eprints}  (\bibinfo{year}{2015}),
  \eprint{1507.03986}.

\bibitem{Gandhi:2015xza}
\bibinfo{author}{\bibfnamefont{R.}~\bibnamefont{Gandhi}},
  \bibinfo{author}{\bibfnamefont{B.}~\bibnamefont{Kayser}},
  \bibinfo{author}{\bibfnamefont{M.}~\bibnamefont{Masud}}, \bibnamefont{and}
  \bibinfo{author}{\bibfnamefont{S.}~\bibnamefont{Prakash}},
  \bibinfo{journal}{JHEP} \textbf{\bibinfo{volume}{11}}, \bibinfo{pages}{039}
  (\bibinfo{year}{2015}), \eprint{1508.06275}.

\bibitem{Datta2004356}
\bibinfo{author}{\bibfnamefont{A.}~\bibnamefont{Datta}},
  \bibinfo{author}{\bibfnamefont{R.}~\bibnamefont{Gandhi}},
  \bibinfo{author}{\bibfnamefont{P.}~\bibnamefont{Mehta}}, \bibnamefont{and}
  \bibinfo{author}{\bibfnamefont{S.~U.} \bibnamefont{Sankar}},
  \bibinfo{journal}{Physics Letters B}
  \textbf{\bibinfo{volume}{597}}(\bibinfo{number}{3–4}), \bibinfo{pages}{356
  } (\bibinfo{year}{2004}).

\bibitem{Chatterjee:2014oda}
\bibinfo{author}{\bibfnamefont{A.}~\bibnamefont{Chatterjee}},
  \bibinfo{author}{\bibfnamefont{R.}~\bibnamefont{Gandhi}}, \bibnamefont{and}
  \bibinfo{author}{\bibfnamefont{J.}~\bibnamefont{Singh}},
  \bibinfo{journal}{JHEP} \textbf{\bibinfo{volume}{1406}}, \bibinfo{pages}{045}
  (\bibinfo{year}{2014}), \eprint{1402.6265}.

\bibitem{Ohlsson:2012kf}
\bibinfo{author}{\bibfnamefont{T.}~\bibnamefont{Ohlsson}},
  \bibinfo{journal}{Rept. Prog. Phys.} \textbf{\bibinfo{volume}{76}},
  \bibinfo{pages}{044201} (\bibinfo{year}{2013}), \eprint{1209.2710}.

\bibitem{Blennow:2005qj}
\bibinfo{author}{\bibfnamefont{M.}~\bibnamefont{Blennow}},
  \bibinfo{author}{\bibfnamefont{T.}~\bibnamefont{Ohlsson}}, \bibnamefont{and}
  \bibinfo{author}{\bibfnamefont{W.}~\bibnamefont{Winter}},
  \bibinfo{journal}{Eur. Phys. J.} \textbf{\bibinfo{volume}{C49}},
  \bibinfo{pages}{1023} (\bibinfo{year}{2007}), \eprint{hep-ph/0508175}.

\bibitem{Adhikari:2012vc}
\bibinfo{author}{\bibfnamefont{R.}~\bibnamefont{Adhikari}},
  \bibinfo{author}{\bibfnamefont{S.}~\bibnamefont{Chakraborty}},
  \bibinfo{author}{\bibfnamefont{A.}~\bibnamefont{Dasgupta}}, \bibnamefont{and}
  \bibinfo{author}{\bibfnamefont{S.}~\bibnamefont{Roy}},
  \bibinfo{journal}{Phys. Rev.} \textbf{\bibinfo{volume}{D86}},
  \bibinfo{pages}{073010} (\bibinfo{year}{2012}), \eprint{1201.3047}.

\bibitem{Barger:2013rha}
\bibinfo{author}{\bibfnamefont{V.}~\bibnamefont{Barger}},
  \bibinfo{author}{\bibfnamefont{A.}~\bibnamefont{Bhattacharya}},
  \bibinfo{author}{\bibfnamefont{A.}~\bibnamefont{Chatterjee}},
  \bibinfo{author}{\bibfnamefont{R.}~\bibnamefont{Gandhi}},
  \bibinfo{author}{\bibfnamefont{D.}~\bibnamefont{Marfatia}}, \bibnamefont{and}
  \bibinfo{author}{\bibfnamefont{M.}~\bibnamefont{Masud}},
  \bibinfo{journal}{Phys. Rev.}
  \textbf{\bibinfo{volume}{D89}}(\bibinfo{number}{1}), \bibinfo{pages}{011302}
  (\bibinfo{year}{2014}), \eprint{1307.2519}.

\bibitem{Barger:2014dfa}
\bibinfo{author}{\bibfnamefont{V.}~\bibnamefont{Barger}},
  \bibinfo{author}{\bibfnamefont{A.}~\bibnamefont{Bhattacharya}},
  \bibinfo{author}{\bibfnamefont{A.}~\bibnamefont{Chatterjee}},
  \bibinfo{author}{\bibfnamefont{R.}~\bibnamefont{Gandhi}},
  \bibinfo{author}{\bibfnamefont{D.}~\bibnamefont{Marfatia}}, \bibnamefont{and}
  \bibinfo{author}{\bibfnamefont{M.}~\bibnamefont{Masud}},
  \bibinfo{journal}{Arxiv eprints}  (\bibinfo{year}{2014}), \eprint{1405.1054}.

\bibitem{Qian:2015waa}
\bibinfo{author}{\bibfnamefont{X.}~\bibnamefont{Qian}} \bibnamefont{and}
  \bibinfo{author}{\bibfnamefont{P.}~\bibnamefont{Vogel}},
  \bibinfo{journal}{Prog. Part. Nucl. Phys.} \textbf{\bibinfo{volume}{83}},
  \bibinfo{pages}{1} (\bibinfo{year}{2015}), \eprint{1505.01891}.

\bibitem{Farzan:2015doa}
\bibinfo{author}{\bibfnamefont{Y.}~\bibnamefont{Farzan}},
  \bibinfo{journal}{Phys. Lett.} \textbf{\bibinfo{volume}{B748}},
  \bibinfo{pages}{311} (\bibinfo{year}{2015}), \eprint{1505.06906}.

\bibitem{Beringer:1900zz}
\bibinfo{author}{\bibfnamefont{J.}~\bibnamefont{Beringer}} \emph{et~al.}
  (\bibinfo{collaboration}{Particle Data Group}), \bibinfo{journal}{Phys. Rev.}
  \textbf{\bibinfo{volume}{D86}}, \bibinfo{pages}{010001}
  (\bibinfo{year}{2012}).

\bibitem{Chatterjee:2014gxa}
\bibinfo{author}{\bibfnamefont{A.}~\bibnamefont{Chatterjee}},
  \bibinfo{author}{\bibfnamefont{P.}~\bibnamefont{Mehta}},
  \bibinfo{author}{\bibfnamefont{D.}~\bibnamefont{Choudhury}},
  \bibnamefont{and} \bibinfo{author}{\bibfnamefont{R.}~\bibnamefont{Gandhi}},
  \bibinfo{journal}{Arxiv eprints}  (\bibinfo{year}{2014}), \eprint{1409.8472}.

\bibitem{Biggio:2009nt}
\bibinfo{author}{\bibfnamefont{C.}~\bibnamefont{Biggio}},
  \bibinfo{author}{\bibfnamefont{M.}~\bibnamefont{Blennow}}, \bibnamefont{and}
  \bibinfo{author}{\bibfnamefont{E.}~\bibnamefont{Fernandez-Martinez}},
  \bibinfo{journal}{JHEP} \textbf{\bibinfo{volume}{0908}}, \bibinfo{pages}{090}
  (\bibinfo{year}{2009}), \eprint{0907.0097}.

\bibitem{Mitsuka:2011ty}
\bibinfo{author}{\bibfnamefont{G.}~\bibnamefont{Mitsuka}} \emph{et~al.}
  (\bibinfo{collaboration}{Super-Kamiokande Collaboration}),
  \bibinfo{journal}{Phys.Rev.} \textbf{\bibinfo{volume}{D84}},
  \bibinfo{pages}{113008} (\bibinfo{year}{2011}), \eprint{1109.1889}.

\bibitem{Adamson:2013ovz}
\bibinfo{author}{\bibfnamefont{P.}~\bibnamefont{Adamson}} \emph{et~al.}
  (\bibinfo{collaboration}{MINOS Collaboration}), \bibinfo{journal}{Phys.Rev.}
  \textbf{\bibinfo{volume}{D88}}(\bibinfo{number}{7}), \bibinfo{pages}{072011}
  (\bibinfo{year}{2013}), \eprint{1303.5314}.

\bibitem{Kopp:2010qt}
\bibinfo{author}{\bibfnamefont{J.}~\bibnamefont{Kopp}},
  \bibinfo{author}{\bibfnamefont{P.~A.} \bibnamefont{Machado}},
  \bibnamefont{and} \bibinfo{author}{\bibfnamefont{S.~J.} \bibnamefont{Parke}},
  \bibinfo{journal}{Phys.Rev.} \textbf{\bibinfo{volume}{D82}},
  \bibinfo{pages}{113002} (\bibinfo{year}{2010}), \eprint{1009.0014}.

\bibitem{Choubey:2015xha}
\bibinfo{author}{\bibfnamefont{S.}~\bibnamefont{Choubey}},
  \bibinfo{author}{\bibfnamefont{A.}~\bibnamefont{Ghosh}},
  \bibinfo{author}{\bibfnamefont{T.}~\bibnamefont{Ohlsson}}, \bibnamefont{and}
  \bibinfo{author}{\bibfnamefont{D.}~\bibnamefont{Tiwari}},
  \bibinfo{journal}{Arxiv eprints}  (\bibinfo{year}{2015}),
  \eprint{1507.02211}.

\bibitem{Huber:2004ka}
\bibinfo{author}{\bibfnamefont{P.}~\bibnamefont{Huber}},
  \bibinfo{author}{\bibfnamefont{M.}~\bibnamefont{Lindner}}, \bibnamefont{and}
  \bibinfo{author}{\bibfnamefont{W.}~\bibnamefont{Winter}},
  \bibinfo{journal}{Comput. Phys. Commun.} \textbf{\bibinfo{volume}{167}},
  \bibinfo{pages}{195} (\bibinfo{year}{2005}), \eprint{hep-ph/0407333}.

\bibitem{Huber:2007ji}
\bibinfo{author}{\bibfnamefont{P.}~\bibnamefont{Huber}},
  \bibinfo{author}{\bibfnamefont{J.}~\bibnamefont{Kopp}},
  \bibinfo{author}{\bibfnamefont{M.}~\bibnamefont{Lindner}},
  \bibinfo{author}{\bibfnamefont{M.}~\bibnamefont{Rolinec}}, \bibnamefont{and}
  \bibinfo{author}{\bibfnamefont{W.}~\bibnamefont{Winter}},
  \bibinfo{journal}{Comput. Phys. Commun.} \textbf{\bibinfo{volume}{177}},
  \bibinfo{pages}{432} (\bibinfo{year}{2007}), \eprint{hep-ph/0701187}.

\bibitem{Dziewonski:1981xy}
\bibinfo{author}{\bibfnamefont{A.~M.} \bibnamefont{Dziewonski}}
  \bibnamefont{and} \bibinfo{author}{\bibfnamefont{D.~L.}
  \bibnamefont{Anderson}}, \bibinfo{journal}{Phys. Earth Planet. Interiors}
  \textbf{\bibinfo{volume}{25}}, \bibinfo{pages}{297} (\bibinfo{year}{1981}).

\bibitem{Freund:2001pn}
\bibinfo{author}{\bibfnamefont{M.}~\bibnamefont{Freund}},
  \bibinfo{journal}{Phys. Rev.} \textbf{\bibinfo{volume}{D64}},
  \bibinfo{pages}{053003} (\bibinfo{year}{2001}), \eprint{hep-ph/0103300}.

\bibitem{Akhmedov:2004ny}
\bibinfo{author}{\bibfnamefont{E.~K.} \bibnamefont{Akhmedov}},
  \bibinfo{author}{\bibfnamefont{R.}~\bibnamefont{Johansson}},
  \bibinfo{author}{\bibfnamefont{M.}~\bibnamefont{Lindner}},
  \bibinfo{author}{\bibfnamefont{T.}~\bibnamefont{Ohlsson}}, \bibnamefont{and}
  \bibinfo{author}{\bibfnamefont{T.}~\bibnamefont{Schwetz}},
  \bibinfo{journal}{JHEP} \textbf{\bibinfo{volume}{0404}}, \bibinfo{pages}{078}
  (\bibinfo{year}{2004}), \eprint{hep-ph/0402175}.

\bibitem{Gandhi:2004bj}
\bibinfo{author}{\bibfnamefont{R.}~\bibnamefont{Gandhi}},
  \bibinfo{author}{\bibfnamefont{P.}~\bibnamefont{Ghoshal}},
  \bibinfo{author}{\bibfnamefont{S.}~\bibnamefont{Goswami}},
  \bibinfo{author}{\bibfnamefont{P.}~\bibnamefont{Mehta}}, \bibnamefont{and}
  \bibinfo{author}{\bibfnamefont{S.~U.} \bibnamefont{Sankar}},
  \bibinfo{journal}{Phys.Rev.} \textbf{\bibinfo{volume}{D73}},
  \bibinfo{pages}{053001} (\bibinfo{year}{2006}), \eprint{hep-ph/0411252}.

\bibitem{Kimura:2002wd}
\bibinfo{author}{\bibfnamefont{K.}~\bibnamefont{Kimura}},
  \bibinfo{author}{\bibfnamefont{A.}~\bibnamefont{Takamura}}, \bibnamefont{and}
  \bibinfo{author}{\bibfnamefont{H.}~\bibnamefont{Yokomakura}},
  \bibinfo{journal}{Phys. Rev.} \textbf{\bibinfo{volume}{D66}},
  \bibinfo{pages}{073005} (\bibinfo{year}{2002}), \eprint{hep-ph/0205295}.

\bibitem{Kimura:2002hb}
\bibinfo{author}{\bibfnamefont{K.}~\bibnamefont{Kimura}},
  \bibinfo{author}{\bibfnamefont{A.}~\bibnamefont{Takamura}}, \bibnamefont{and}
  \bibinfo{author}{\bibfnamefont{H.}~\bibnamefont{Yokomakura}},
  \bibinfo{journal}{Phys. Lett.} \textbf{\bibinfo{volume}{B537}},
  \bibinfo{pages}{86} (\bibinfo{year}{2002}), \eprint{hep-ph/0203099}.

\bibitem{Kimura:2006jj}
\bibinfo{author}{\bibfnamefont{K.}~\bibnamefont{Kimura}},
  \bibinfo{author}{\bibfnamefont{A.}~\bibnamefont{Takamura}}, \bibnamefont{and}
  \bibinfo{author}{\bibfnamefont{T.}~\bibnamefont{Yoshikawa}},
  \bibinfo{journal}{Phys. Lett.} \textbf{\bibinfo{volume}{B640}},
  \bibinfo{pages}{32} (\bibinfo{year}{2006}).

\bibitem{Asano:2011nj}
\bibinfo{author}{\bibfnamefont{K.}~\bibnamefont{Asano}} \bibnamefont{and}
  \bibinfo{author}{\bibfnamefont{H.}~\bibnamefont{Minakata}},
  \bibinfo{journal}{JHEP} \textbf{\bibinfo{volume}{1106}}, \bibinfo{pages}{022}
  (\bibinfo{year}{2011}), \eprint{1103.4387}.

\bibitem{Ohlsson:2013ip}
\bibinfo{author}{\bibfnamefont{T.}~\bibnamefont{Ohlsson}},
  \bibinfo{author}{\bibfnamefont{H.}~\bibnamefont{Zhang}}, \bibnamefont{and}
  \bibinfo{author}{\bibfnamefont{S.}~\bibnamefont{Zhou}},
  \bibinfo{journal}{Phys. Rev.}
  \textbf{\bibinfo{volume}{D87}}(\bibinfo{number}{5}), \bibinfo{pages}{053006}
  (\bibinfo{year}{2013}), \eprint{1301.4333}.

\bibitem{Kikuchi:2008vq}
\bibinfo{author}{\bibfnamefont{T.}~\bibnamefont{Kikuchi}},
  \bibinfo{author}{\bibfnamefont{H.}~\bibnamefont{Minakata}}, \bibnamefont{and}
  \bibinfo{author}{\bibfnamefont{S.}~\bibnamefont{Uchinami}},
  \bibinfo{journal}{JHEP} \textbf{\bibinfo{volume}{0903}}, \bibinfo{pages}{114}
  (\bibinfo{year}{2009}), \eprint{0809.3312}.

\bibitem{Masud:2016bvp}
\bibinfo{author}{\bibfnamefont{M.}~\bibnamefont{Masud}} \bibnamefont{and}
  \bibinfo{author}{\bibfnamefont{P.}~\bibnamefont{Mehta}}
  (\bibinfo{year}{2016}), \eprint{1603.01380}.

\bibitem{deGouvea:2015ndi}
\bibinfo{author}{\bibfnamefont{A.}~\bibnamefont{de~Gouva}} \bibnamefont{and}
  \bibinfo{author}{\bibfnamefont{K.~J.} \bibnamefont{Kelly}},
  \bibinfo{journal}{Nucl. Phys.} \textbf{\bibinfo{volume}{B908}},
  \bibinfo{pages}{318} (\bibinfo{year}{2016}), \eprint{1511.05562}.

\bibitem{Coloma:2015kiu}
\bibinfo{author}{\bibfnamefont{P.}~\bibnamefont{Coloma}},
  \bibinfo{journal}{JHEP} \textbf{\bibinfo{volume}{03}}, \bibinfo{pages}{016}
  (\bibinfo{year}{2016}), \eprint{1511.06357}.

\bibitem{Liao:2016hsa}
\bibinfo{author}{\bibfnamefont{J.}~\bibnamefont{Liao}},
  \bibinfo{author}{\bibfnamefont{D.}~\bibnamefont{Marfatia}}, \bibnamefont{and}
  \bibinfo{author}{\bibfnamefont{K.}~\bibnamefont{Whisnant}},
  \bibinfo{journal}{Phys. Rev.}
  \textbf{\bibinfo{volume}{D93}}(\bibinfo{number}{9}), \bibinfo{pages}{093016}
  (\bibinfo{year}{2016}), \eprint{1601.00927}.

\bibitem{Forero:2016cmb}
\bibinfo{author}{\bibfnamefont{D.~V.} \bibnamefont{Forero}} \bibnamefont{and}
  \bibinfo{author}{\bibfnamefont{P.}~\bibnamefont{Huber}}
  (\bibinfo{year}{2016}), \eprint{1601.03736}.

\end{thebibliography}

\end{document}